\documentclass[twocolumn,amsmath,amssymb,prc,superscriptaddress,nofootinbib]{revtex4-1}

\usepackage{graphicx}

\usepackage{dcolumn}
\usepackage{bm}
\usepackage{longtable}
\usepackage{siunitx}
\usepackage{booktabs}
\usepackage{etoolbox}
\usepackage{xcolor}
\robustify\itshape

\AtBeginDocument{
\heavyrulewidth=.08em
\lightrulewidth=.05em
\cmidrulewidth=.03em
\belowrulesep=.65ex
\belowbottomsep=0pt
\aboverulesep=.4ex
\abovetopsep=0pt
\cmidrulesep=\doublerulesep
\cmidrulekern=.5em
\defaultaddspace=.5em
}

\newcolumntype{.}{D{.}{.}{1}}
\newcolumntype{X}{D{X}{X}{1}}

\begin{document}

\newcommand{\reaction}[6]{\nuc{#1}{#2}(#3,#4)\/\nuc{#5}{#6}}
\newcommand{\nuc}[2]{\ensuremath{^{#1}}#2}
\newcommand{\Kpp}[0]{\ensuremath{^{39}}K\ensuremath{+}p }
\newcommand{\Erlab}[1]{\ensuremath{E_{r}^{\text{lab}} = #1}~keV}
\newcommand{\Ercm}[1]{\ensuremath{E_{r}^{\text{c.m.}} = #1}~keV}
\newcommand{\Ercmb}[1]{\ensuremath{\mathbf{E_{r}^{\textbf{c.m.}} = #1}}~\textbf{keV}}
\newcommand{\Ex}[1]{\ensuremath{E_{x} = #1}~keV}
\newcommand{\Exb}[1]{\ensuremath{\mathbf{E_{x} = #1}}~\textbf{keV}}
\renewcommand{\pg}[0]{\reaction{39}{K}{p}{$\gamma$}{40}{Ca} }
\newcommand{\pa}[0]{\reaction{39}{K}{p}{$\alpha$}{36}{Ar} }
\newcommand{\Jpi}[2]{\ensuremath{J^{\pi} = #1^{#2}} }
\newcommand{\Jpib}[2]{\ensuremath{\mathbf{J^{\pi} = #1^{#2}}} }
\def\sun{\odot}

\title{Reaction rates for the \pg reaction}

\author{R. Longland} \affiliation{Department of Physics, North
  Carolina State University, Raleigh, NC 27695, USA}
\affiliation{Triangle Universities Nuclear Laboratory, Durham, NC
  27708, USA}

\author{J. Dermigny} 
\affiliation{Department of Physics and Astronomy, University of North
  Carolina at Chapel Hill, Chapel Hill, NC 27599, USA}
\affiliation{Triangle Universities Nuclear Laboratory, Durham, NC
  27708, USA}
\author{C. Marshall} \affiliation{Department of Physics, North
  Carolina State University, Raleigh, NC 27695, USA}
\affiliation{Triangle Universities Nuclear Laboratory, Durham, NC
  27708, USA}


\begin{abstract}
  The magnesium-potassium anti-correlation observed in globular
  cluster NGC2419 can be explained by nuclear burning of hydrogen in
  hot environments. The exact site of this nuclear burning is, as yet,
  unknown. In order to constrain the sites responsible for this
  anti-correlation, the nuclear reactions involved must be well
  understood.  The \Kpp reactions are one such pair of
  reactions. Here, we report a new evaluation of the \pg reaction rate
  by taking into account ambiguities and measurement uncertainties in
  the nuclear data. The uncertainty in the \pg reaction rate is larger
  than previously assumed, and its influence on nucleosynthesis models
  is demonstrated. We find the \pg reaction cross
  section should be the focus of future experimental study to help
  constrain models aimed at explaining the magnesium-potassium
  anti-correlation in globular clusters.
\end{abstract}


\maketitle


\section{Introduction}
\label{sec:intro}


The alkali element potassium is synthesized in several stellar
environments. It is predominately produced in a combination of
hydrostatic and explosive oxygen burning~\cite{WW_2002}, conditions
found only in highly-evolved massive stars during the pre-supernova
phase and the ensuing explosion. However, models of galactic chemical
evolution, which are based on the nucleosynthetic yields of
supernovae, so far severely under-predict the observed potassium
abundance in our
galaxy~\cite{Timmes_1995,Goswami_2000,Romano_2010,Prantzos2018}. Potassium
is also synthesized in smaller quantities in high-temperature hydrogen
burning environments, which are believed to be important in explaining
elemental abundance signatures in globular clusters. Of particular
interest is NGC 2419~\cite{Mucciarelli_2012}, where it was found that
a significant fraction of its member stars ($\approx40\%$) are highly
enriched in elemental potassium.  Additionally, there is a strong
anticorrelation observed between potassium and magnesium abundances,
reminiscent of the ubiquitous Na-O and Mg-Al anticorrelations found
much more commonly in clusters (see Ref. \cite{Gratton_2012} and
references therein). Though the main reason for these discrepancies
has not been established, a more accurate description of potassium
synthesis will be helpful in this area. To achieve this, the potassium
destruction reactions \Kpp are crucial.


Iliadis \textit{et al.} \cite{Iliadis2016} explored the astrophysical
conditions that could be responsible for isotopic correlations in NGC
2419. Their method featured a Monte Carlo nucleosynthesis network that
included all known uncertainties in the thermonuclear reaction
rates. They obtained a range of stellar temperatures and densities
that quantitatively reproduced all of the elemental abundances
measured in the potassium-rich stars. Later, Ref. \cite{Dermigny_2017}
extended that method by including a sensitivity study of the nuclear
reaction rates. They found several reactions whose rates need to be
better constrained in order to more accurately identify an
astrophysical site responsible for the anomalies. The majority of
these pertain to the synthesis and destruction of \nuc{39}{K}. The
rate of one of these reactions, \pg, was based on preliminary
calculations, so it is important to reinvestigate the $^{39}$K $+$ p
reactions based on a full evaluation of the nuclear physics input.


In this paper, we calculate the rate of the \pg reaction using all
available experimental information. The \pa rate was found to not
significantly influence final abundances in the stellar environments
of interest here, so we leave its evaluation to future work. In
Sec. \ref{sec:formalism}, a brief overview of the reaction rate
formalism is presented along with the Monte Carlo method used to
calculate uncertainties on the rate given experimental uncertainties
on the cross sections. In Sec.~\ref{sec:rates}, details of the
experimental information are presented. The Monte Carlo rates using
that information are computed in Sec.~\ref{sec:rates-results} and
compared to previous reaction rate calculations. Astrophysical
implications of these rates as they pertain to nucleosynthesis in
globular clusters are presented in Sec.~\ref{sec:astro-results}, and all is
summarized in Sec.~\ref{sec:conclusions}.

\section{Reaction Rate Formalism}
\label{sec:formalism}
\subsection{Thermonuclear Reaction Rates}
\label{sec:rates-formalism}

In a stellar plasma, the rate of a nuclear reaction per particle pair
is given by
\begin{equation}
  \label{eq:rates-reacrate}
  \langle\sigma v\rangle = \sqrt{\frac{8}{\pi \mu}}
  \frac{1}{(kT)^{3/2}}\int_{0}^{\infty}E\sigma(E)e^{-E/kT} dE.
\end{equation}
Here $\mu$ is the reduced mass of the reacting particles, $k$ is the
Boltzmann constant, $T$ is the temperature of the plasma, and
$\sigma(E)$ is the energy-dependent cross section of the reaction. For
a slowly-varying cross section or one consisting of multiple broad,
overlapping, or interfering resonances, the integral in
Eqn.~\eqref{eq:rates-reacrate} must be solved numerically. However,
for isolated, narrow resonances, it can be replaced by an incoherent sum over
their individual contributions:
\begin{equation}
  \label{eq:rates-narrowrate}
  \langle\sigma v\rangle = \left(\frac{2\pi}{\mu
      kT}\right)^{3/2}\hbar^2 \sum_i \omega\gamma_i e^{-E_{r,i}/kT},
\end{equation}
where the resonance strength, $\omega \gamma_i$ for resonance $i$ at
energy $E_{r,i}$ is defined by
\begin{equation}
  \label{eq:rates-ResStrength}
  \omega \gamma = \omega \frac{\Gamma_a \Gamma_b}{\Gamma}.
\end{equation}
$\Gamma_a$ and $\Gamma_b$ are the entrance and exit particle partial
widths, and $\Gamma$ is the total width given by the sum of widths
over all open reaction channels. $\omega$ is the spin factor. The
particle partial width for channel $c$, $\Gamma_c$, can be written as
\begin{equation}
  \label{eq:rates-GammaDefinition}
  \Gamma_{c} = 2\, P_c(\!E_r\!)\, \gamma_c^2,
\end{equation}
where $P_c(\!E_r\!)$ is the penetration factor at the resonance energy and
$\gamma_c^2$ is the energy-independent reduced width. The reduced
width can be calculated by
\begin{equation}
  \label{eq:S-reduced-width-relation1}
  \gamma^2_{c} = \frac{\hbar^2}{2 \mu R}\, C^2 S\, \phi^2_{R}.
\end{equation}
$R$ is the channel radius given by $R=R_0 ( A_a^{1/3} + A_b^{1/3}
)$.
$\phi^2_{R}$ is the single-particle radial wave function at the
channel radius, which can be calculated theoretically
\cite{ILI97,BEL07}.

The \pg reaction proceeds through the compound nucleus \nuc{40}{Ca} at
a high excitation energy ($S_{p}=8328.437 (21)$\ keV~
\citep{AME12}). The average level density of \nuc{40}{Ca} at these
excitation energies is about 60 MeV$^{-1}$, corresponding to an
average level spacing of about 20 keV. Of these levels, fewer will
exhibit an appreciable proton width and contribute significantly to
the reaction rate, as will become apparent in
Sec.~\ref{sec:rates-results}. At the low proton energies relevant for
astrophysics, the high Coulomb barrier renders the proton width in
Eqn.~\eqref{eq:rates-GammaDefinition} to be small. Thus, the cross
section can be considered to be dominated by narrow
resonances and the reaction rate is calculated using
Eqn.~\eqref{eq:rates-narrowrate}. Interference effects are expected to
average to a negligible contribution, and the non-resonant part of the
cross section can be neglected.

\subsection{Monte Carlo Reaction Rates}
\label{sec:rates-montecarlo}

The resonance strengths, partials widths, and resonance energies used
to calculate reaction rates in Eqs. \eqref{eq:rates-narrowrate} and
\eqref{eq:rates-ResStrength} are obtained from experimental
information, theoretical estimates, or are unknown. They must,
therefore, carry some associated uncertainty whose probability density
distribution varies depending on the source of that uncertainty. The
uncertainty in these parameters results in an uncertainty in the
reaction rate. Traditionally, crude estimates of ``upper'' and
``lower'' limits of the reaction rate have been computed by
considering which parameter possibilities can be combined to maximize
or minimize the reaction rate. This was the case, for example, in the
NACRE evaluation of reaction rates~\cite{ANG99}. However, these
methods define unphysical bounds on a reaction rate, whose uncertainty
distribution should be continuous. Other, more sophisticated methods
have also been employed by attempting full uncertainty propagation
techniques~\cite{THO99,ILI01}. However, those techniques could not
account for parameters with large uncertainties, numerically
integrated cross sections, or partial widths for which only upper
limits are known.

Here, we utilize a Monte Carlo uncertainty propagation method. Using
this method, probability density distributions can be fully defined
for all uncertain input parameters in a reaction rate
calculation. These methods are described in detail in
Refs.~\cite{LON10,ILI10a}. In summary, the central limit theorem
suggests that \textit{measured} partial widths, resonance strengths,
or cross sections should have uncertainties dictated by log-normal
probability density distributions whose location and shape parameters
are calculated from the expectation value and variance of the
experimental data. \textit{Unmeasured}, or so-called ``upper limit''
partial widths have uncertainties dictated by the Porter-Thomas
probability density distribution~\cite{Porter1956}. Resonance
energies, on the other hand, are dictated by normal, or Gaussian,
probability density distributions. The strategy of Monte Carlo
uncertainty propagation is straight forward: (i) a random sample of
each uncertain parameter is obtained using its probability density
distribution; (ii) a reaction rate sample is calculated using these
sample parameters in Eqs.~\eqref{eq:rates-reacrate} or
\eqref{eq:rates-narrowrate}; (iii) a new set of parameter samples is
generated as in step (i). These steps are repeated, taking care to
correctly account for energy effects in partial width calculations,
many times (typically 3,000-10,000 times depending on available
computing power and the complexity of the cross section). Following
this procedure, an ensemble of reaction rate samples is obtained for
each temperature, which can be summarized in meaningful statistics. In
Ref.~\cite{LON10}, we found that the log-normal shape parameters,
$\mu$ and $\sigma$, can well summarize the probability density
distribution of Monte Carlo reaction rates. The code \texttt{RatesMC}\
\citep{LON10} was used to perform the Monte Carlo sampling and to
analyse the probability densities of the total reaction rates.

The log-normal probability density distribution used for resonance
strengths and partial widths is defined by
\begin{equation}
  \label{eq:rates-lognormal}
  f(x) = \frac{1}{\sigma \sqrt{2 \pi}} \frac{1}{x} e^{-(\ln x -
    \mu)^2/(2 \sigma^2)},
\end{equation}
where $x$ represents the resonance strength (or partial
    width) defined in units of eV here.
The log-normal parameters, $\mu$ and $\sigma$ do not represent the
mean and standard deviation as with a normal distribution, but rather
the mean and standard deviation of $\ln x$. Note that a lognormal
distribution is only defined for positive values of $x$. The
parameters $\mu$ and $\sigma$ are related to the expectation value,
$E[x]$, and variance, $V[x]$, by
\begin{equation}
  \label{eq:rates-lognormpars}
  E[x] = e^{(2\mu + \sigma^2)/2}, \qquad V[x] = e^{(2\mu +
    \sigma^2)} \left[ e^{\sigma^2} - 1 \right],
\end{equation}
where $E[x]$ and $V[x]$ are defined here in units of eV and eV$^2$,
respectively. The values $\ln(E[x])$ and $\sqrt V[x]$ can be
associated with the central value and uncertainty in resonance
strengths or partial widths that are commonly reported in the
literature. Often, rather than a variance, reaction cross sections can
be reported with a \textit{factor uncertainty} (f.u.). For a
log-normal probability density distribution, the recommended value
then refers to the geometric mean of the reaction rate, i.e., the
median reaction rate. The factor uncertainties refer to multiplicative
factors describing an uncertainty. The lognormal parameters are
calculated using $\mu = \ln(\text{rec.})$ and
$\sigma = \ln(\text{f.u.})$. The expectation value and variance can
then be calculated using Eqns.~\eqref{eq:rates-lognormpars}. For
example, consider a fictional resonance with an estimated strength of
$\omega \gamma = 3$~eV with a factor uncertainty of 2. We must
interpret this reported value as the \textit{geometric} mean of the
resonance strength with high and low values at $3 \cdot 2 = 6$~eV and
$3/2 = 1.5$~eV, respectively (recall that \textit{high} and
\textit{low} here do not refer to hard limits). The lognormal
parameters are $\mu = \ln(3) = 1.10$ and $\sigma = \ln(2) = 0.69$.
Using the identities in Eqn.~\eqref{eq:rates-lognormpars}, the
expectation value and variance of the resonance strength are found to
be $E[\omega \gamma] = 3.8$~eV and $V[\omega \gamma] = 3.0$~eV,
respectively. These values are used as input to our \texttt{RatesMC}
input to ensure correct lognormal conversion.

For many reactions, the reaction rate at low temperatures is expected
to be dominated by resonances whose partial widths are unknown. They
are governed only by an upper limit, either experimental or
theoretical. A thorough discussion of this issue is available
elsewhere (see, for example, Refs. \cite{LON10,LON12,Pogrebnyak2013}),
and will be summarized here. Particle partial widths depend on
overlaps between the entrance channel (\nuc{39}{K} + p) and the final
state in the compound nucleus. Compound nuclear states can be defined
by way of nuclear matrix elements, which often contain contributions
from many different parts of configuration space whose signs are
randomly distributed. The central limit theorem predicts that the
probability density function of the transition amplitude will tend
toward a Gaussian distribution centered on zero. The reduced width,
defined as the square of this transition amplitude, will thus be
distributed according to a chi-squared distribution with one degree of
freedom. For a particle channel, this probability density function for
the single particle reduced width is given by
\begin{equation}
  \label{eq:PT}
  f(\theta) = \frac{c}{\sqrt{\theta^2}}e^{-\theta^2/(2 \hat{\theta}^2)}
\end{equation}
where $c$ is a normalization constant, $\theta^2$ is the dimensionless
reduced width, and $\hat{\theta}^2$ is the local mean value of the
dimensionless reduced width, which has been investigated in
Ref.~\cite{Pogrebnyak2013}. This distribution, also known as the
Porter-Thomas distribution \cite{POR56}, is well established
theoretically~\cite{WEI09}. It also gives us a physically motivated
probability density distribution from which to sample unknown or
upper-limit particle partial widths. The mean dimensionless reduced
width used here is obtained from Tab. II in
Ref. \cite{Pogrebnyak2013}. For the case of measured upper limits, the
distribution in Eqn.~\eqref{eq:PT} is truncated at the upper-limit
value. Note that this is an approximation: upper limit measurements
themselves contain a probability distribution, not a sharp cut-off as
this strategy currently assumes. More sophisticated strategies such as
matching the 95th percentile, for example, should be investigated.

Often, resonance strength or partial width measurements are normalized
to reference resonances. If this is the case, their properties are
correlated.  Longland~\cite{Longland2017} investigated the effect of
these correlations on Monte Carlo reaction rates and recommended
techniques for accounting for them. In short, it was found that
defining a single correlation parameter, $\rho_i$ for each resonance,
$i$, was sufficient for accounting for correlation between resonance
strengths and partial widths. By assuming that any normalized
resonance strength must necessarily have larger uncertainties than the
reference resonance, correlation parameters can be defined by
\begin{equation}
  \label{eq:rho}
  \rho_i = \frac{\sigma_r}{\omega
    \gamma_r} \frac{\omega \gamma_i}{\sigma_i} = \frac{f.u._r}{f.u._i},
\end{equation}
where $\omega \gamma$ and $\sigma$ are the resonance strengths and
their associated uncertainties, respectively. The subscript $r$ refers
to the reference resonance. Note that given this definition, the
correlation parameter cannot be larger than 1. A correlated Monte
Carlo resonance strength sample, $\omega \gamma_{i,j}$ can then be
calculated using
\begin{equation}
  \label{eq:res-strength-sample}
  \omega \gamma_{i,j} = \omega \gamma_{i,\textrm{rec.}} \, (f.u.)^{y'_{i,j}},
\end{equation}
where the recommended value is denoted by $\omega
\gamma_{i,\textrm{rec.}}$, its factor uncertainty is $(f.u.)$, and
$y'_{i,j}$ is a correlated, normally distributed random sample
calculated by
\begin{equation}
  y_{i,j}' = \rho_i x_{j} + \sqrt{1-\rho_i^2}\, \, y_{i,j}. \label{eq:yp}
\end{equation}
Here, $x_j$ and $y_{i,j}$ are uncorrelated, normally distributed
random variables with a mean of 0 and standard deviation of 1. The
correlated resonance strengths calculated using
Eqn.~\eqref{eq:res-strength-sample} can then be used in the standard
Monte Carlo procedure. 

Following this procedure of constructing probability density
distributions for uncertain input parameters and calculating the
reaction rates using the Monte Carlo technique described above, an
ensemble of reaction rates is obtained. These are also expected, in
most cases, to follow a lognormal distribution whose location and
shape parameters can be calculated using
Eqns.~\eqref{eq:rates-lognormpars}. However, this is an
approximation. In some cases, particularly those for which the
resonance strength is dominated by upper limits of resonances, the
lognormal assumption is not valid.

A useful measure of the applicability of a lognormal approximation to
the actual sampled distribution is provided by the Anderson-Darling
statistic, which is calculated from
\begin{equation}
  \label{eq:rates-AD}
  t_{AD} = -n - \sum_{i=1}^n\frac{2i-1}{n}(\ln F(y_i) +
  \ln\left[1-F(y_{n+1-i})\right]
\end{equation}
where $n$ is the number of samples, $y_i$ are the sampled reaction
rates at a given temperature (arranged in ascending order), and $F$ is
the cumulative distribution of a standard normal function (i.e., a
Gaussian centred at zero). An A-D value greater than unity indicates a
deviation from a lognormal distribution. However, it was found by
\textcite{LON10} that the rate distribution does not \emph{visibly}
deviate from lognormal until A-D exceeds $t_{AD} \approx 30$.  The A-D
statistic is presented in Tab.~\ref{tab:pgRate} along with the
reaction rates at each temperature in order to provide a reference to
the reader.

\section{The \Kpp Reactions}
\label{sec:rates}

\subsection{General Aspects}
\label{sec:general-aspects}

The \Kpp reactions proceed through excited states in \nuc{40}{Ca} with
Q-values of $Q_{(p,\gamma)} = 8.328437 (21)$~MeV for the \pg reaction
and $Q_{(p,\alpha)} = 1.288675 (27)$~MeV for the \pa reaction\
\cite{AME12}. The ground-state spins of \nuc{39}{K}, \nuc{40}{Ca}, and
\nuc{36}{Ar} are \Jpi{3/2}{+}, \Jpi{0}{+}, and \Jpi{0}{+},
respectively. Note that since the first excited state in \nuc{36}{Ar}
is at \Ex{1970} with \Jpi{2}{+}, the \pa reaction can only proceed
through natural parity states in \nuc{40}{Ca} at energies below about
\Ercm{1000}. This is true for the relevant proton energies
corresponding to the temperature range $T=200-300$ MK as identified by
Ref.~\cite{Dermigny2017}. Those energies are between \Ercm{200} and
\Ercm{600}, corresponding to an excitation energy region in
\nuc{40}{Ca} of \Ex{8500 - 9000}.

Several studies have investigated the cross section of the \pg
reaction by way of resonance strength determination. These are listed
in Tab.~\ref{tab:direct-resonance-pg}. Note that for the \pa reaction,
only one investigation has been performed for incident proton energies
below about \Erlab{3000}~\cite{deMeijer1970}. Indirect measurements
have also been performed, which provide useful supplementary
information. Proton widths in the excitation energy region of interest
have been inferred by proton transfer reactions: through
($^3$He,d)~\cite{Erskine1966,Seth1967,Forster1970,Cage1971} and
(d,n)~\cite{Fuchs1969}. Additionally, $\alpha$-particle widths in this
excitation energy range have been inferred through the
\reaction{36}{Ar}{$^{6}$Li}{d}{40}{Ca} reaction~\cite{Yamaya1994}.

\subsection{Resonance Strengths}
\label{sec:resonance-strengths}

\begin{table*}
  \def\arraystretch{1.3}
  \centering
    \begin{tabular}{cc|l}
      \hline \hline
      Reference                & Reaction Studied & Comments                                                                \\ \hline
      \textcite{Leenhouts1966} & \pg              & Relative $\omega \gamma$ between \Ercm{763} and \Ercm{2747}                  \\
      \textcite{Cheng1981}     & \pg              & Absolute $\omega \gamma$ for \Ercm{1102}, \Ercm{1310}, and \Ercm{1992}. \\
                               &                  & Relative $\omega \gamma$ between \Ercm{763} and \Ercm{2747}             \\
      \textcite{Kikstra1990}   & \pg              & Relative $\omega \gamma$ between \Ercm{606} and \Ercm{2838}             \\
      \hline \hline
    \end{tabular}
  \caption{\label{tab:direct-resonance-pg}Summary of direct resonance
    strength measurements in the energy region of interest.}
\end{table*}

\begin{figure}
  \centering
  \includegraphics[width=0.45\textwidth]{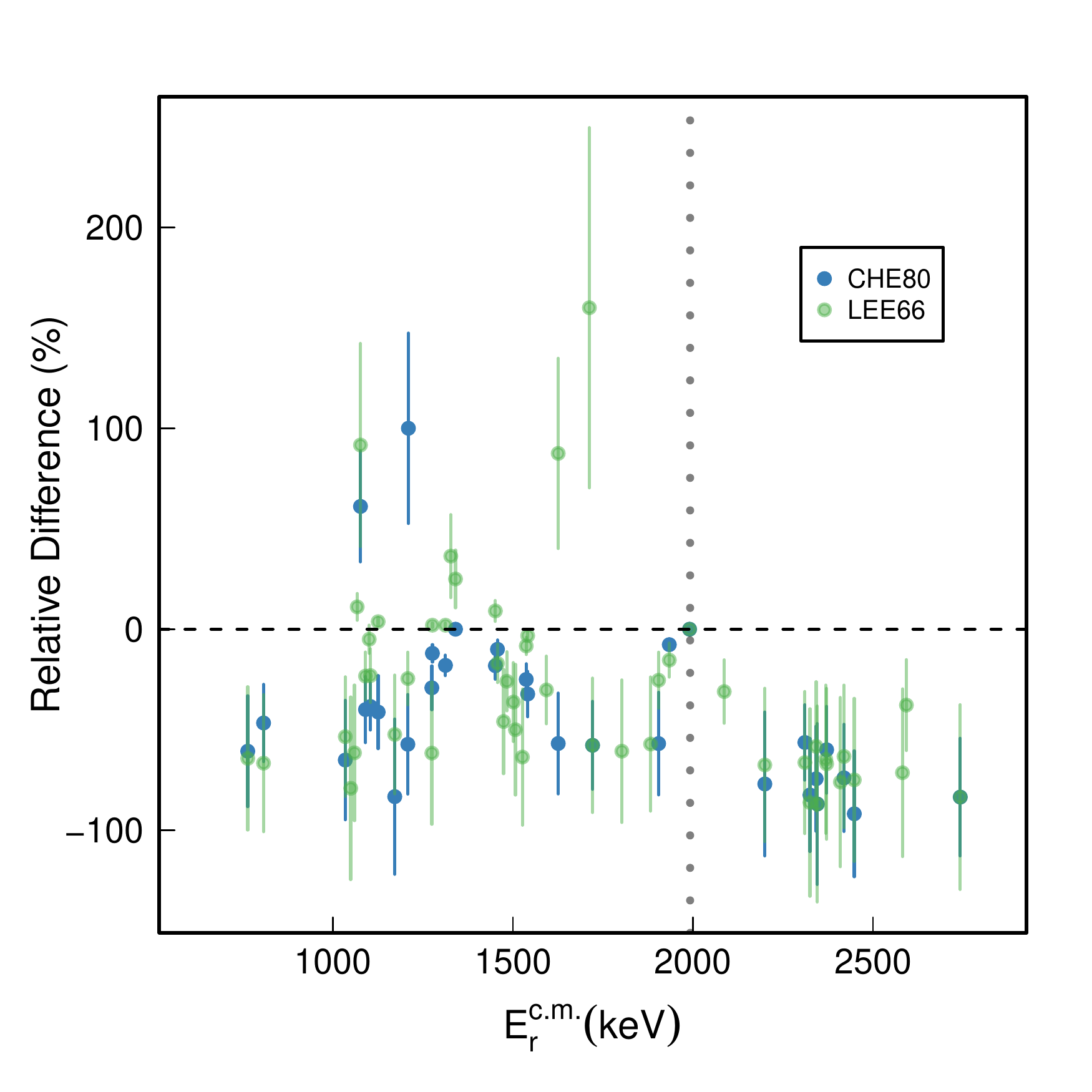}
  \caption{(color online) Comparison to experimental resonance
    strengths compared to the data of
    Ref. \cite{Kikstra1990}. Uncertainties in the points include the
    uncertainties reported in Ref. \cite{Kikstra1990}. Clearly,
    although there is agreement between resonance strengths at the
    normalization resonance at \Ercm{1992} (shown as a vertical dashed
    line), the agreement is poor at most other energies. There appears
    to be an overall systematic shift to higher resonance strengths in
    Ref. \cite{Kikstra1990}.}
  \label{fig:wgCompare-unnorm}
\end{figure}

Resonance strengths for the \pg reaction have been measured directly
(see also Tab.~\ref{tab:direct-resonance-pg}) by Leenhouts \textit{et
  al.}~\cite{Leenhouts1966}, Cheng \textit{et al.}~\cite{Cheng1981},
and Kikstra \textit{et al.}~\cite{Kikstra1990}. The latter of these
normalized their results to the \Ercm{1990} (\Erlab{2043}) resonance
measured absolutely by Ref. \cite{Cheng1981}. The evaluation in
Ref. \cite{Chen2017} adopted resonance strengths from the most recent
measurement (Ref. \cite{Kikstra1990}). Earlier measurements should
still be valid, though, and close inspection reveals a systematic
shift of resonance strengths in Ref. \cite{Kikstra1990}. This is shown
in Fig.~\ref{fig:wgCompare-unnorm}, where resonance strengths measured
by Ref.~\cite{Kikstra1990} and Refs.~\cite{Leenhouts1966,Cheng1981}
are compared as a function of resonance energy after normalization to
the common \Ercm{1990} resonance strength reported by
Ref.~\cite{Cheng1981}. The strengths are shown relative to those of
Ref.~\cite{Kikstra1990}, which lies along the line at
zero. Uncertainties in the points include the uncertainties reported
in Ref. \cite{Kikstra1990}. The normalization point at \Ercm{1990} is
denoted by a vertical dotted line. Clearly, large differences exist
between the measured resonance strengths that are outside their
uncertainties\footnote{Kikstra \textit{et al.}~\cite{Kikstra1990} also
  comment on this disagreement and postulate that the thick-target
  method used by Ref.~\cite{Cheng1981} could be the culprit. However,
  upon inspection we do not find this argument convincing.}. In order
to fully describe, in probabilistic terms, our confidence in resonance
strength determinations for astrophysical purposes, a more careful
evaluation of these measurements is necessary.

\begin{figure}
  \centering
  \includegraphics[width=0.45\textwidth]{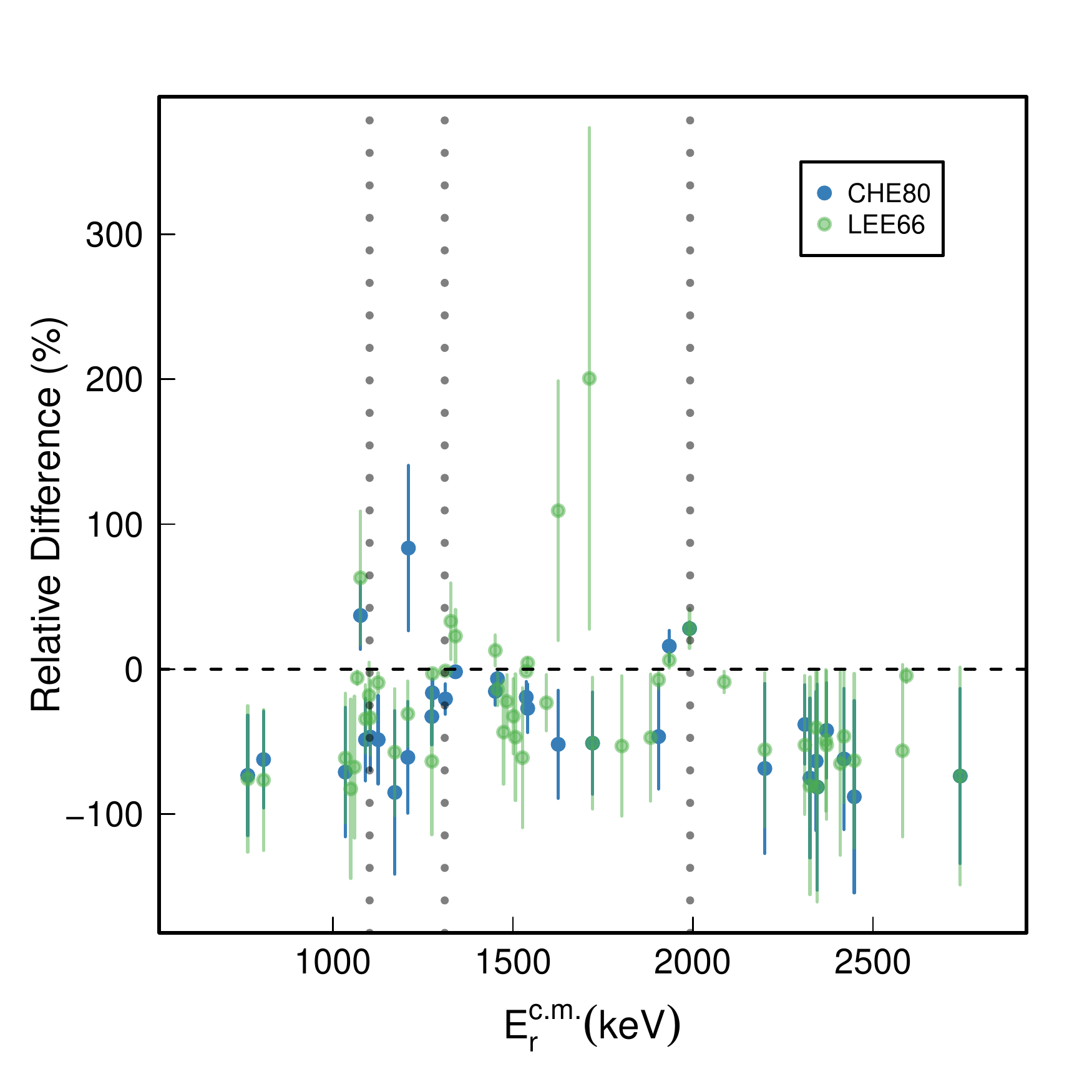}
  \caption{(color online) Comparison to experimental resonance
    strengths compared to the data of
    Ref. \cite{Kikstra1990}. Uncertainties in the points include the
    uncertainties reported in Ref. \cite{Kikstra1990} and additional
    uncertainty introduced by our re-normalization process. In this
    case, the Kikstra resonance strengths have been reduced by a
    factor of 1.28 as described in the text. The new normalization
    points are shown by three vertical dashed lines. Although there is
    still a systematic disagreement between data sets, it is now
    within the 2$\sigma$ uncertainties.}
  \label{fig:wgCompare-renorm}
\end{figure}

Our procedure is to first correct the data from
Ref.~\cite{Kikstra1990} for target stopping powers, which are an
energy-dependent quantity not accounted for in their
analysis. Secondly, we recognize that the reference resonance used for
normalization comes from Ref.~\cite{Cheng1981}, where, in fact,
\textit{three} absolute measurements were performed at \Ercm{1104},
\Ercm{1312}, and \Ercm{1990}. If all three of those absolute resonance
strengths (taking their uncertainties into account) are used to
normalize the strengths in Ref.~\cite{Kikstra1990}, better agreement
is obtained. This results in a \textit{reduction} in the
Ref.~\cite{Kikstra1990} resonance strengths by a factor of 1.28. The
result of these two corrections is shown in
Fig. \ref{fig:wgCompare-renorm}. Poor agreement between measurements
is still present.

\begin{figure*}
  \centering
  \includegraphics[width=0.8\textwidth]{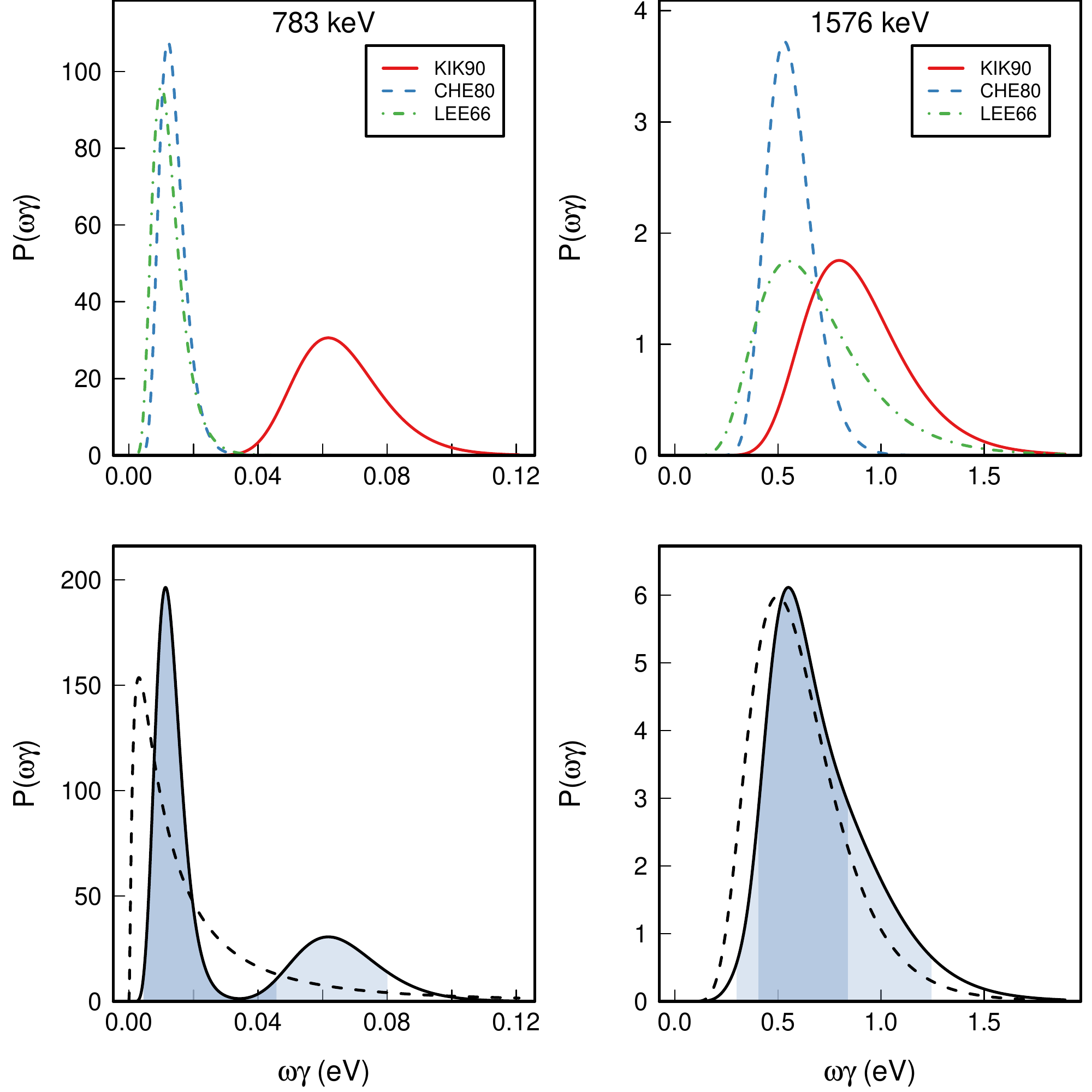}
  \caption{(color online) Example probability density functions for
    two experimentally measured resonances (at \Ercm{763} and
    \Ercm{1537}) as and reported in Ref.~\cite{Leenhouts1966}
    ("LEE66"), Ref.~\cite{Cheng1981} ("CHE80"), and
    Ref.~\cite{Kikstra1990} ("KIK90"), respectively. By summing the
    probability density functions incoherently, the HDPI Bayesian
    method can be utilized to summarize our confidence in the
    experimental results. The dark and light shaded regions correspond
    to the 68\% and 95\% confidence intervals. The
        resulting log-normal probability density approximation is
        shown in the lower panels as a black, dashed line. See text
    for more detail.}
  \label{fig:PDF-Example}
\end{figure*}

To account for this remaining disagreement between measured resonance
strengths in our calculations, we consider the probability density
distributions corresponding to the reported uncertainties in the
measurements. After investigating a number of descriptive statistics,
we found that calculating the Highest Density Posterior Interval
(HDPI)~\cite{BoxBook} for these distributions provides a good
representation of the data. An example of this procedure can be seen
in Fig.~\ref{fig:PDF-Example}. First, probability density
distributions for each reported resonance strength are
constructed. These probabilities are expected to follow a lognormal
distribution as discussed in Sec.~\ref{sec:rates-montecarlo}. The
individual probability density distributions are shown in the top
panels of Fig.~\ref{fig:PDF-Example} for two resonance strengths at
\Ercm{763} and \Ercm{1537}. The reported strengths for the \Ercm{763}
resonance contain some disagreement in Ref.~\cite{Kikstra1990}, while
for the \Ercm{1537} resonance, there is good agreement between the
reported values. Once constructed, the probability distributions are
summed incoherently, as shown by the solid line in the lower panels. A
bi-modal distribution is clearly evident for the \Ercm{763}
resonance. Finally, the HDPI is computed, which consists of finding
the smallest range of values that contain a given integrated
probability. This procedure naturally includes the mode (highest
point) of the distribution. Figure~\ref{fig:PDF-Example} shows two
such regions in dark and light blue (grey in print version) for a 68\%
and 95\% coverage, respectively.

Once a 68\% uncertainty range has been computed, it must be converted
into a representation that is used as input in the \texttt{RatesMC}
Monte Carlo reaction rate code. Input to the code is assumed to be
expectation values and variances. To compute those values, we first
assume that the HDPI describes the 1-$\sigma$ uncertainties of a
lognormal distribution. Given that assumption, the lognormal location
and shape parameters, $\mu$ and $\sigma$ can be calculated from the
low ($x_{\text{low}}$) and high ($x_{\text{high}}$) interval values
using
\begin{equation}
  \label{eq:musig}
  \mu = \ln \sqrt{x_{\text{low}} x_{\text{high}}}, \qquad \qquad \sigma
= \ln \sqrt{\frac{x_{\text{high}}}{x_{\text{low}}}}.
\end{equation}
The expectation value and variance can be calculated using
Eqn.~\eqref{eq:rates-lognormpars}.

It should be stressed that this methodology is equivalent to
calculating a weighted mean when measurements are in agreement, but
also accounts for unknown systematic effects. These expectation and
variance values are calculated for every resonance measured by
Refs. \cite{Leenhouts1966}, \cite{Cheng1981}, and \cite{Kikstra1990},
and are summarized in Tab.~\ref{tab:direct-pg}.



\subsection{Indirectly Measured Information}
\label{sec:indirect}

Although direct measurements of resonance strengths have been
performed, the resonances have all been above the effective stellar
burning range of \Ercm{200-600}. To supplement the direct data,
partial width and spin-parity assignments from other, indirect,
measurements can be used. Proton widths in the astrophysically
important range have been determined though the (d,n) reaction by
Fuchs \textit{et al.}~\cite{Fuchs1969}, and the ($^3$He,d) reaction by
Erskine \textit{et al.}~\cite{Erskine1966}, Seth \textit{et
  al.}~\cite{Seth1967}, and Cage \textit{et al.}~\cite{Cage1971}. Of
these, Ref.~\cite{Cage1971} employed a more advanced finite range
modified Distorted Wave Born Approximation, so we consider it to
supersede the other studies here. However, it should be noted that
their results are in good agreement with the (d,n) measurement of
Ref. \cite{Fuchs1969}. 

Alpha-particle widths are obtained from
\reaction{36}{Ar}{$^{6}$Li}{d}{40}{Ca}~\cite{Yamaya1994}. However, the
energy levels and spin-parities extracted in that work do not clearly
align with the levels observed in proton transfer. Thus they cannot be
applied reliably to calculating reaction rates for the \pg reaction
and are ignored for that reaction rate calculation (they are expected
to have a negligible effect on the \pg rate at lower
temperatures). 

For resonances with no known proton partial width, upper limits are
used as discussed in Sec.~\ref{sec:rates-montecarlo}. We assume upper
limits on spectroscopic factors of S=1, so the quantity $C^2S$ in
Eqn.~\eqref{eq:S-reduced-width-relation1} is assumed not to exceed
$C^2S=0.5$. For the calculations presented here, we only consider
upper limit resonances below the lowest directly measured resonance at
\Ercm{606}. Above this value, we assume that the rate is dominated by
resonances that have been measured.

In order to calculate proton reduced widths given the expected
Porter-Thomas probability density function described in
Sec.~\ref{sec:rates-montecarlo}, the spin-parity of the state should
be constrained. In the excitation energy region between the proton
threshold at \Ex{8328} and the lowest directly measured resonance at
\Ex{8935}, 21 states have unknown proton widths. Their spins have been
determined through decay scheme analysis and ($^6$Li,d) alpha-particle
transfer measurements. We adopt the values evaluated in
Ref. \cite{Chen2017}.

\begin{table*}

    \sisetup{
      table-alignment=center,
      retain-zero-exponent=true,
    }
    \begin{tabular}{
      S[table-format=7.2(2)]
      S[table-format=6.2(2)]
      c
      S[table-format=5.2e2]
      S[table-format=6.2(2)e2]
      }
      \toprule \toprule 
      {E$_{x}$ (keV)} & {E$_{r}^{\textrm{c.m.}}$ (keV)} & {J$^{\pi}$} & {$\Gamma_{p,\mathrm{UL}}$} & {$\Gamma_{\alpha}$} \\ \midrule 
        8338.0 (3)    & 9.6   (3)                     & (2+,3,4)    & 2.47E-72                   &                     \\
        8358.9	 (3)  & 30.5   (6)                    & (0,1,2)-    & 5.88E-36                   &                     \\    
        8364	(5)   & 36     (5)                    & (3- to 7-)  & 4.29E-33                   &                     \\
        8373.94 (15)  & 45.50  (15)                   & 4+          & 3.05E-29                   & 4.5 (25)e-9         \\
        8424.81  (11) & 96.37  (11)                   & 2-          & 2.94E-16                   &                     \\
        8439.0	  (5) & 110.6  (5)                    & 0+          & 1.23E-15                   &                     \\
        8484.02  (13) & 155.58 (13)                   & (1-,2-,3-)  & 3.88E-10                   &                     \\
        8540	(4)   & 212    (4)                    & 1,2+        & 3.16E-07                   &                     \\
        8551.1	(7)   & 222.7  (7)                    & 5-          & 6.00E-10                   &                     \\
        8578.80 (9)   & 250.36 (9)                    & 2+          & 8.16E-06                   &                     \\
        8587 (2)      & 259    (2)                    & (2+,3)      & 1.47E-05                   &                     \\
        8633 (6)      & 305    (6)                    & 2+          & 2.57E-04                   & 6.3 (31)e-6         \\
        8665.3	(8)   & 336.9  (8)                    & 1-          & 3.66E-04                   &                     \\
        8678.29 (10)  & 349.85 (10)                   & 4+          & 5.86E-05                   &                     \\
        8701 (1)      & 372.6  (10)                   & (6-)        & 5.11E-12                   &                     \\
        8717 (8)      & 389    (8)                    & 0           & 1.18E-02                   &                     \\
        8748.22 (9)   & 419.78 (9)                    & 2+          & 3.62E-02                   &                     \\
        8764.18 (6)   & 435.74 (6)                    & 3-          & 1.72E-02                   &                     \\
        8810 (7)      & 482    (7)                    & 2+          & 2.37E-01                   & 8.6 (43)e-5         \\
        8850.6	(9)   & 522.2  (9)                    & 6-,7-,8-    & 1.43E-05                   &                     \\
        8909.0 (9)    & 580.6  (9)                    & 0           & 2.48E+00                   &                     \\
        8934.81 (7)   & 606.37 (7)                    & 2+          & 4.14E+00                   & 5.5 (23)e-4         \\
        8935.8	(9)   & 607.4  (9)                    & (7+)        & 1.09E-04                   &                     \\
        8938.4	(9)   & 610.0  (9)                    & 0+          & 1.28E-01                   &                     \\
        8978 (6)      & 650    (6)                    & 5+,6+,7+    & 1.17E-02                   &                     \\
        9162.1	(11)  & 833.7	(11)                  & 2+          & 1.25E+02                   & 2.5 (13)e-2         \\
        9246.0	(12)  & 917.6	(12)                  & (7)-        & 2.61E-04                   & 2.6 (13)e-5         \\
        9362.54	(6)   & 1034.10	(6)                   & 3-          & 2.94E+02                   & 2.2(11)e-2          \\
        9499.9	(15)  & 1171.5	(15)                  & 2+          & 2.54E+03                   & 1.07 (53)e-1        \\
        9668.71	(8)   & 1340.27	(8)                   & 3-          & 2.47E+03                   & 5.1(25)e-1          \\
        9869.3	(4)   & 1540.9	(4)                   & 1+,2+       & 1.86E+04                   & 2.1 (11)e0          \\
        9954.00	(9)   & 1625.56	(9)                   & 4+          & 1.52E+03                   & 7.5(38)e-1          \\
        10058.0	(3)   & 1729.6	(3)                   & (1-,2+)     & 3.84E+04                   & 9.3(46)e0           \\
        10130.70 (19) & 1802.26	(19)                  & 5-          & 2.58E+02                   & 4.0(20)e-1          \\
        10318.8	(4)   & 1990.36	(40)                  & 8+          & 1.96E-02                   & 2.3(12)e-2          \\
        \hline \hline 
      \end{tabular} 
      \caption{\label{tab:an_upperlims}Properties of Unobserved
        Resonances in \pg and \pa. For these resonances, only upper limits of
        the resonance strength can be derived.}
\end{table*}

\subsection{Information on Specific \nuc{40}{Ca} Levels}
\label{sec:levels}

Three states below the lowest measured resonance at \Ercm{606} in \pg
have inferred proton partial widths from proton transfer
reactions. Four have $\alpha$-particle partial widths assigned from
($^6$Li,d) measurements. Here, we will address these states
individually. 

\begin{description}
\item[\Exb{8373.94} (\Ercmb{46}, \Jpib{4}{+})] The spin-parity
  assignment of this state is based on (see Ref.~\cite{Chen2017} and
  references therein) inelastic $\alpha$-particle scattering,
  \reaction{42}{Ca}{p}{t}{40}{Ca}, and
  \reaction{36}{Ar}{$^6$Li}{d}{30}{Ca}. The latter study also assigned
  an $\alpha$-particle spectroscopic factor of S$_{\alpha}=0.043$.
\item[\Exb{8424.81} (\Ercmb{96}, \Jpib{2}{-})] This state cannot
  contribute to the \pa reaction rate due to its unnatural parity. Its
  spin-parity comes from \reaction{41}{Ca}{$^3$He}{$\alpha$}{40}{Ca},
  inelastic proton scattering, \reaction{39}{K}{d}{n}{40}{Ca}, and
  \reaction{39}{K}{$^3$He}{d}{40}{Ca}. A proton spectroscopic factor
  of S$_{p}=0.56$ has been determined~\cite{Cage1971}. 
\item[\Exb{8551.1} (\Ercmb{223}, \Jpib{5}{-})] The spin-parity of the
  \Ex{8551} state is well established~\cite{Chen2017}. A large proton
  spectroscopic factor has been obtained in multiple
  \reaction{39}{K}{$^3$He}{d}{40}{Ca} measurements. Here, we adopt the
  coupled-channel result of S$_{p}=0.84$ from
  Ref.~\cite{Cage1971}. The $\alpha$-particle spectroscopic factor
  from Ref.~\cite{Yamaya1994} is S$_{\alpha}=0.043$.
\item[\Exb{8633} (\Ercmb{305}, \Jpib{2}{+})] A \Jpi{2}{+} state at
  \Ex{8600} was observed in \reaction{36}{Ar}{$^6$Li}{d}{30}{Ca} by
  Ref. \cite{Yamaya1994}. We assign this to the state observed in
  inelastic proton scattering at \Ex{8633}. However, their resolution
  and statistics could lead to an incorrect assignment. There are two
  other \Jpi{2}{+} in the vicinity at \Ex{8587} and \Ex{8578}. This
  state could also correspond to the \Jpi{5}{-} state at \Ex{8551}. We
  note that no \Jpi{0}{+} state was observed in this region by
  Ref.~\cite{Yamaya1994} as expected from the average level spacing in
  Ref. \cite{Fortune1975}. Careful inspection of the angular
  distributions presented in Refs.~\cite{Yamaya1994} does not rule out
  a \Jpi{0}{+} assignment for this state. Clearly, higher resolution
  studies are required to precisely identify $\alpha$-decaying states
  using $\alpha$-particle transfer.
\item[\Exb{8665} (\Ercmb{337}, \Jpib{1}{-})] This state has been
  populated by inelastic proton scattering and the
  \reaction{39}{K}{d}{n}{40}{Ca} reaction, but \textit{not} by the
  \reaction{39}{K}{$^3$He}{d}{40}{Ca} reaction, where it is obscured
  by background from the \reaction{16}{O}{$^3$He}{d}{17}{F} reaction
  in the target. The proton spectroscopic factor is S$_p=0.19$.
\item[\Exb{8810} (\Ercmb{482}, \Jpib{2}{+})] A \Jpi{2}{+} state was
  populated by the \reaction{36}{Ar}{$^6$Li}{d}{30}{Ca} in
  Ref. \cite{Yamaya1994} at \Ex{8780}. The closest known \Jpi{2}{+}
  state to this is at \Ex{8810} and has been observed in inelastic
  $\alpha$-particle scattering \cite{VanDerBorg1981} and inelastic
  proton scattering. It has an inferred $\alpha$-particle
  spectroscopic factor of S$_{\alpha}=0.11$.
\end{description}

\section{Reaction Rates for \pg}
\label{sec:rates-results}

Using the information detailed in Sec.~\ref{sec:rates}, rates for the
\pg reaction are calculated using the Monte Carlo method outlined in
Sec.~\ref{sec:formalism}. Those rates are shown in
Tab.~\ref{tab:pgRate}. 

\begin{figure} 
  \begin{center}
    \includegraphics[width=0.45\textwidth]{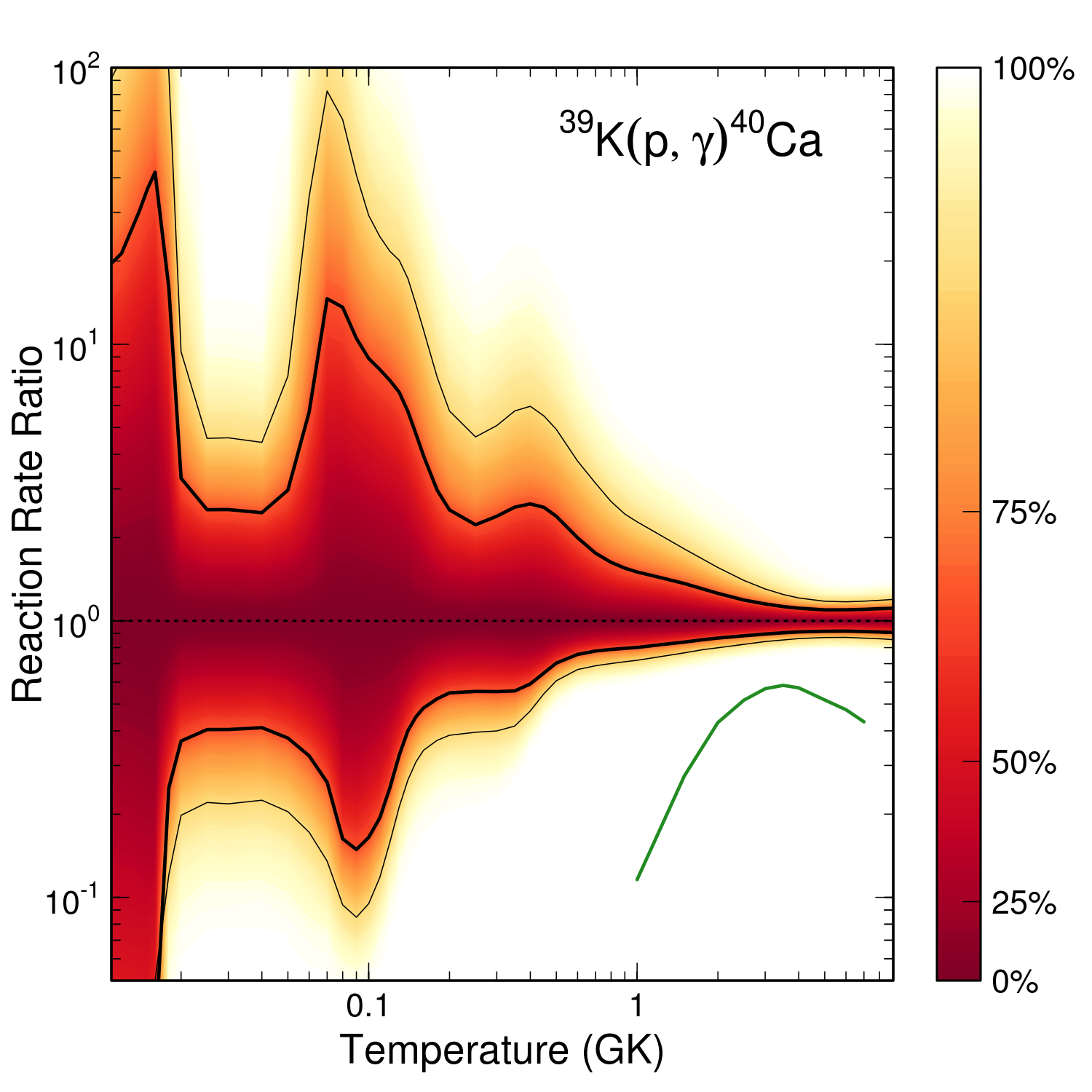} 
    \caption[\pg rate distributions]{\label{fig:GraphContour-pg}(Color
      online) Reaction rate probability densities for the \pg
      reaction. The reaction rate has been normalized to the median,
      recommended rate. Hence the thick and thin lines correspond to
      the 1- and 2-$\sigma$ uncertainty bands. The color scale
      highlights that the rate probability distribution at each
      temperature is continuous with no absolute upper or lower
      limit. The solid green (grey) line represents the most recent
      calculation of Ref. \cite{Cheng1981}. }
  \end{center}
\end{figure}

The \pg reaction rate is shown as a contour plot in
Fig.~\ref{fig:GraphContour-pg}. The contour is normalized to the
recommended (median) rate at each temperature, so this figure serves
to illustrate the temperature-dependent uncertainty in the reaction
rate. Darker (red) colors represent higher probability
values close to the recommended rate, with lighter
(yellow online) colors showing lower probability
values. Clearly there is no sharp cut-off of the reaction rate
probability distribution. For convenience, the 68\% and 95\%
uncertainty bands are shown in thick and thin black lines,
respectively. At 100~MK, for example, the 95\% uncertainties span
three orders of magnitude. The reaction rate has previously been
computed in Ref.~\cite{Cheng1981} for T=1-9~GK. Their results are
clearly lower than our calculated rates, as shown by the solid green
(grey in print version) line in
Fig.~\ref{fig:GraphContour-pg}. This disagreement arises from new
experimental information in Ref.~\cite{Kikstra1990}.

\begin{figure} 
  \begin{center}
    \includegraphics[width=0.45\textwidth]{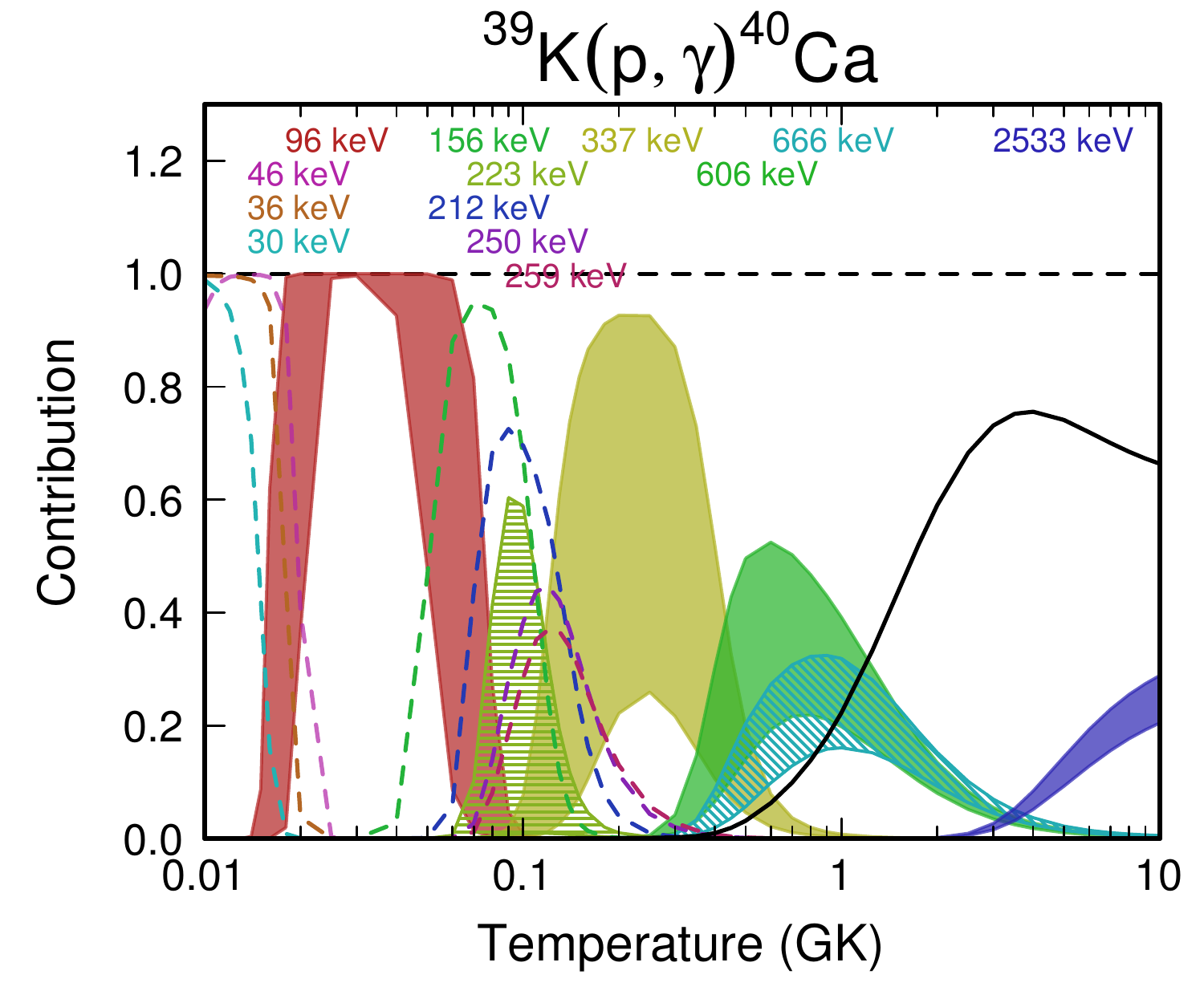}
    \caption[\pg Contributions]{\label{fig:GraphContribution-pg}(Color
      online) Fractional resonance contributions to the \pg reaction
      rate as a function of energy. A value of unity indicates that a
      particular resonance is responsible for 100\% of the reaction
      rate at that temperature. The finite width of resonance
      contributions reflects their uncertainty calculated with the
      Monte Carlo method. To improve clarity, upper limit resonance
      contributions are shown by dashed lines at their 84\% ``high''
      contribution value. The solid line at high temperatures
      represents the summed contribution of all resonances that,
      individually, contribute less than 20\% to the total reaction
      rate.}
  \end{center} 
\end{figure}

To identify the resonances dominating the reaction rate at a
particular energy, a contribution plot for the \pg reaction is shown
in Fig. \ref{fig:GraphContribution-pg}. Inspection of that figure
indicates that the large rate uncertainties at 100~MK arise from the
resonance at \Ercm{223} which has experimentally determined proton and
$\alpha$-particle widths, as well as upper limit resonances at
\Ercm{212}, \Ercm{250}, and \Ercm{259}. The \Ercm{337} resonance
dominates the reaction rate between about 100~MK and 500~MK. Clearly
these resonances should be the focus of any future experimental
investigation.


\begin{table*} 
  \sisetup{
    table-alignment=center,
    retain-zero-exponent=true,
    input-symbols = {()},
    explicit-sign,
    table-space-text-pre = (,
    table-space-text-post = ),
    table-align-text-pre = false,
    detect-all = true
  }
  \resizebox{!}{90mm}{ 
    \begin{tabular}{
      S
      S[table-format=4.2e4]
      S[table-format=3.2e4]
      S[table-format=3.2e4]
      S[table-format=3.3e2]
      S[table-format=3.2e2]
      S[table-format=3.2e2]
      } 
      \toprule \toprule 
      {T (GK)} & {Low rate} & {Median rate} & {High rate} & {lognormal $\mu$} & {lognormal $\sigma$} & {A-D}    \\ \midrule
       0.010   & 5.19e-48   & 4.23e-46      & 7.93e-45    & -1.052e+02        & 3.97e+00             & 1.25e+02 \\
        0.011  & 2.44e-46   & 1.79e-44      & 3.51e-43    & -1.014e+02        & 3.99e+00             & 1.26e+02 \\
        0.012  & 5.73e-45   & 4.47e-43      & 9.51e-42    & -9.819e+01        & 4.02e+00             & 1.14e+02 \\
        0.013  & 7.85e-44   & 6.72e-42      & 1.72e-40    & -9.540e+01        & 4.03e+00             & 9.22e+01 \\
        0.014  & 7.52e-43   & 7.24e-41      & 2.20e-39    & -9.293e+01        & 3.94e+00             & 6.31e+01 \\
        0.015  & 6.35e-42   & 5.72e-40      & 2.09e-38    & -9.065e+01        & 3.69e+00             & 3.99e+01 \\
        0.016  & 1.07e-40   & 3.67e-39      & 1.54e-37    & -8.841e+01        & 3.25e+00             & 6.49e+01 \\
        0.018  & 6.77e-38   & 2.72e-37      & 4.37e-36    & -8.369e+01        & 2.07e+00             & 1.95e+02 \\
        0.020  & 1.95e-35   & 5.30e-35      & 1.74e-34    & -7.881e+01        & 1.19e+00             & 4.59e+01 \\
        0.025  & 7.76e-31   & 1.92e-30      & 4.84e-30    & -6.842e+01        & 9.21e-01             & 1.68e-01 \\
        0.030  & 1.02e-27   & 2.52e-27      & 6.36e-27    & -6.124e+01        & 9.22e-01             & 1.57e-01 \\
        0.040  & 8.00e-24   & 1.95e-23      & 4.79e-23    & -5.229e+01        & 9.04e-01             & 1.33e-01 \\
        0.050  & 1.80e-21   & 4.79e-21      & 1.42e-20    & -4.671e+01        & 1.10e+00             & 3.12e+01 \\
        0.060  & 7.52e-20   & 2.32e-19      & 1.32e-18    & -4.262e+01        & 1.59e+00             & 1.73e+02 \\
        0.070  & 1.57e-18   & 6.02e-18      & 8.79e-17    & -3.914e+01        & 1.99e+00             & 1.87e+02 \\
        0.080  & 3.30e-17   & 2.03e-16      & 2.76e-15    & -3.581e+01        & 2.08e+00             & 1.29e+02 \\
        0.090  & 7.13e-16   & 4.78e-15      & 5.02e-14    & -3.277e+01        & 1.96e+00             & 9.26e+01 \\
        0.100  & 1.11e-14   & 6.72e-14      & 5.97e-13    & -3.016e+01        & 1.83e+00             & 8.36e+01 \\
        0.110  & 1.23e-13   & 6.34e-13      & 5.14e-12    & -2.790e+01        & 1.71e+00             & 9.29e+01 \\
        0.120  & 1.11e-12   & 4.43e-12      & 3.28e-11    & -2.590e+01        & 1.58e+00             & 1.16e+02 \\
        0.130  & 8.25e-12   & 2.52e-11      & 1.69e-10    & -2.410e+01        & 1.44e+00             & 1.46e+02 \\
        0.140  & 5.05e-11   & 1.26e-10      & 7.20e-10    & -2.248e+01        & 1.31e+00             & 1.73e+02 \\
        0.150  & 2.51e-10   & 5.59e-10      & 2.65e-09    & -2.101e+01        & 1.19e+00             & 1.89e+02 \\
        0.160  & 1.05e-09   & 2.18e-09      & 8.61e-09    & -1.969e+01        & 1.08e+00             & 1.90e+02 \\
        0.180  & 1.21e-08   & 2.31e-08      & 6.84e-08    & -1.740e+01        & 9.19e-01             & 1.61e+02 \\
        0.200  & 8.72e-08   & 1.59e-07      & 4.00e-07    & -1.551e+01        & 8.19e-01             & 1.25e+02 \\
        0.250  & 3.09e-06   & 5.56e-06      & 1.24e-05    & -1.199e+01        & 7.43e-01             & 8.15e+01 \\
        0.300  & 3.49e-05   & 6.29e-05      & 1.50e-04    & -9.542e+00        & 7.75e-01             & 1.03e+02 \\
        0.350  & 2.23e-04   & 4.00e-04      & 1.03e-03    & -7.664e+00        & 8.05e-01             & 1.49e+02 \\
        0.400  & 1.09e-03   & 1.85e-03      & 4.90e-03    & -6.100e+00        & 7.88e-01             & 2.19e+02 \\
        0.450  & 4.56e-03   & 7.04e-03      & 1.81e-02    & -4.744e+00        & 7.32e-01             & 3.00e+02 \\
        0.500  & 1.61e-02   & 2.28e-02      & 5.45e-02    & -3.564e+00        & 6.64e-01             & 3.59e+02 \\
        0.600  & 1.20e-01   & 1.59e-01      & 3.17e-01    & -1.665e+00        & 5.46e-01             & 3.92e+02 \\
        0.700  & 5.32e-01   & 6.85e-01      & 1.20e+00    & -2.407e-01        & 4.65e-01             & 3.64e+02 \\
        0.800  & 1.65e+00   & 2.10e+00      & 3.42e+00    & 8.539e-01         & 4.14e-01             & 3.21e+02 \\
        0.900  & 4.03e+00   & 5.07e+00      & 7.86e+00    & 1.720e+00         & 3.79e-01             & 2.85e+02 \\
        1.000  & 8.29e+00   & 1.03e+01      & 1.55e+01    & 2.423e+00         & 3.54e-01             & 2.60e+02 \\
        1.250  & 3.18e+01   & 3.86e+01      & 5.51e+01    & 3.728e+00         & 3.08e-01             & 2.30e+02 \\
        1.500  & 8.22e+01   & 9.82e+01      & 1.34e+02    & 4.648e+00         & 2.72e-01             & 2.18e+02 \\
        1.750  & 1.70e+02   & 1.99e+02      & 2.60e+02    & 5.345e+00         & 2.40e-01             & 2.05e+02 \\
        2.000  & 3.03e+02   & 3.50e+02      & 4.40e+02    & 5.899e+00         & 2.13e-01             & 1.87e+02 \\
        2.500  & 7.18e+02   & 8.13e+02      & 9.66e+02    & 6.728e+00         & 1.71e-01             & 1.41e+02 \\
        3.000  & 1.34e+03   & 1.50e+03      & 1.72e+03    & 7.329e+00         & 1.42e-01             & 9.97e+01 \\
        3.500  & 2.17e+03   & 2.40e+03      & 2.70e+03    & 7.793e+00         & 1.22e-01             & 6.77e+01 \\
        4.000  & 3.19e+03   & 3.50e+03      & 3.88e+03    & 8.168e+00         & 1.07e-01             & 4.28e+01 \\
        5.000  & 5.67e+03   & 6.19e+03      & 6.78e+03    & 8.734e+00         & 9.35e-02             & 1.41e+01 \\
        6.000  & 8.47e+03   & 9.22e+03      & 1.01e+04    & 9.133e+00         & 9.15e-02             & 6.29e+00 \\
        7.000  & 1.12e+04   & 1.23e+04      & 1.35e+04    & 9.420e+00         & 9.43e-02             & 5.26e+00 \\
        8.000  & 1.38e+04   & 1.51e+04      & 1.67e+04    & 9.628e+00         & 9.82e-02             & 5.59e+00 \\
        9.000  & 1.60e+04   & 1.76e+04      & 1.96e+04    & 9.781e+00         & 1.02e-01             & 6.07e0   \\
        10.000 & 1.78e+04   & 1.97e+04      & 2.20e+04    & 9.892e+00         & 1.05e-01             & 4.63e0   \\
      \bottomrule   \bottomrule 
    \end{tabular}
  }
  \caption{\label{tab:pgRate} Monte Carlo reaction rates for the
    \pg reaction. Shown are the
    low, median, and high rates, corresponding to the 16th, 50th,
    and 84th percentiles of the Monte Carlo probability density
    distributions. Also shown are the parameters ($\mu$ and
    $\sigma$) of the lognormal approximation to the actual Monte
    Carlo probability density, as well as the Anderson-Darling
    statistic (A-D). See Ref.\ \cite{LON10} for details. }
\end{table*}

\section{Astrophysical Implications}

\label{sec:astro-results}

To investigate the astrophysical implications of these reaction rates,
we performed a nucleosynthesis calculation based on the findings of
Dermigny and Iliadis~\cite{Dermigny2017}. Using a single-zone
nucleosynthesis model, they found the temperature and density
conditions that reproduced the observed abundances of all elements up
to vanadium in the globular cluster NGC 2419. Their findings indicate
that the observations could be matched between $\mathrm{T} = 100$~MK,
$\rho = 10^8 \, \mathrm{g/cm}^3$ and $\mathrm{T} = 200$~MK,
$\rho = 10^{-4} \, \mathrm{g/cm}^3$.  From these bounds, a
representative environment with temperature and density of
$\mathrm{T} = 170$~MK and $\rho = 100 \, \mathrm{g/cm}^3$ was selected
to test the updated rates and their uncertainties.  Using initial
abundances from Ref. \cite{Iliadis2016}, the network was run until the
mass fraction of hydrogen fell to $X(H)_f = 0.5$.

Holding these parameters constant, a Monte Carlo study of the reaction
rate uncertainties was carried out using STARLIB v6.2~\cite{SAL13} .
The STARLIB library\footnote{Current version of STARLIB is available
  at https://github.com/Starlib/Rate-Library} incorporates the
probabilistic rate formalism described in
Sec. \ref{sec:rates-montecarlo} by giving the median rate and factor
uncertainty ($e^{\sigma}$) over a grid of temperatures.  Following the
methods of Ref. \cite{Iliadis2015}, these parameters can be used to
draw samples from the rates according to:

\begin{equation}
  \label{eq:sample}
  x(T) = x_{med} \times \mathrm{f.u.}^{p}, 
\end{equation}

where $p$ is the so called rate variation factor.
During each run of the network, a value, $p_i$, is drawn
from a standard normal distribution for each nuclei. Therefore, rates whose uncertainty
strongly influences the production of potassium will have a 
correlation between $p_i$ and the final abundances of potassium.
Following the suggestions of Ref. \cite{Iliadis2015}, the degree of correlation
is measured using Spearman's rank correlation coefficient.





\begin{figure*}
  \centering
  \includegraphics[width=0.7\textwidth]{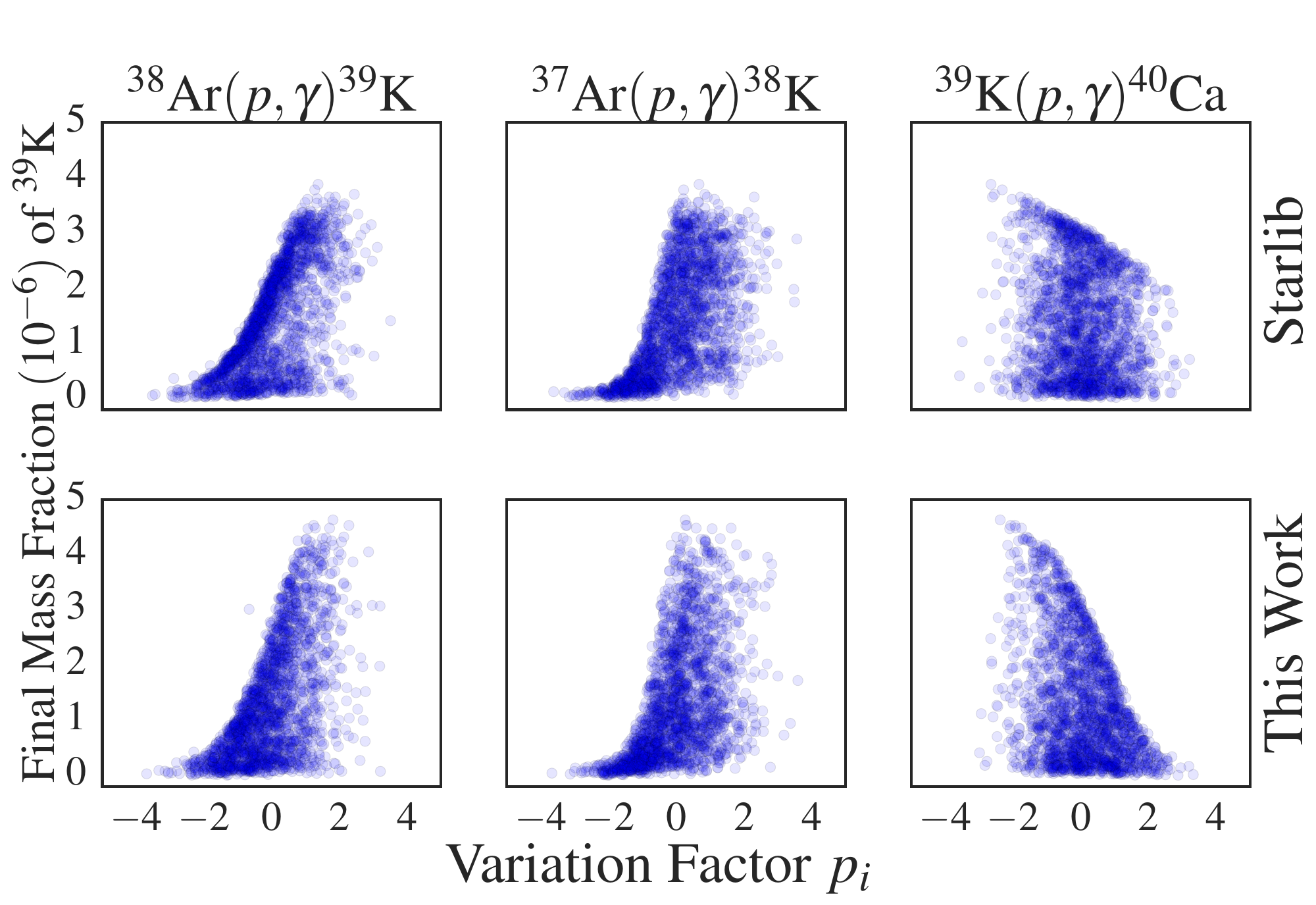}
  \caption{\label{fig:MCCorrelations}(Color online) Monte Carlo
    nucleosynthesis results at $\mathrm{T} = 170$~MK and
    $\rho = 100 \, \mathrm{g/cm}^3$. The upper row represents results
    for the STARLIB reaction rates, while the lower row contains those
    obtained here by fully evaluating all experimental information for
    the \pg and \pa reaction. The increased correlation and importance
    of \pg on potassium production is seen in the two right most
    plots.}
\end{figure*}

The network was run 2,000 times with all rates being simultaneously
sampled from Eq.~\ref{eq:sample}.  A comparison was made by
substituting the reevaluated \pg reaction rate and its reverse rate
into STARLIB.  The correlations between the final \nuc{39}{K} mass
fraction and each reaction in the network were analyzed.  It was found
that only 3 reactions in the network have an appreciable correlation
with the final \nuc{39}{K} abundance. As seen in
Fig.~\ref{fig:MCCorrelations}, the original STARLIB rates display
large correlations for both \reaction{38}{Ar}{p}{$\gamma$}{39}{K} and
\reaction{37}{Ar}{p}{$\gamma$}{38}{K}, but the dependence on \pg is
noticeably weaker.  However, for the new rates all three of these
reactions display clear, strong correlations, and the production of
\nuc{39}{K} is critically sensitive to the rate of \pg.

\begin{figure}
  \centering
  \includegraphics[width=0.45\textwidth]{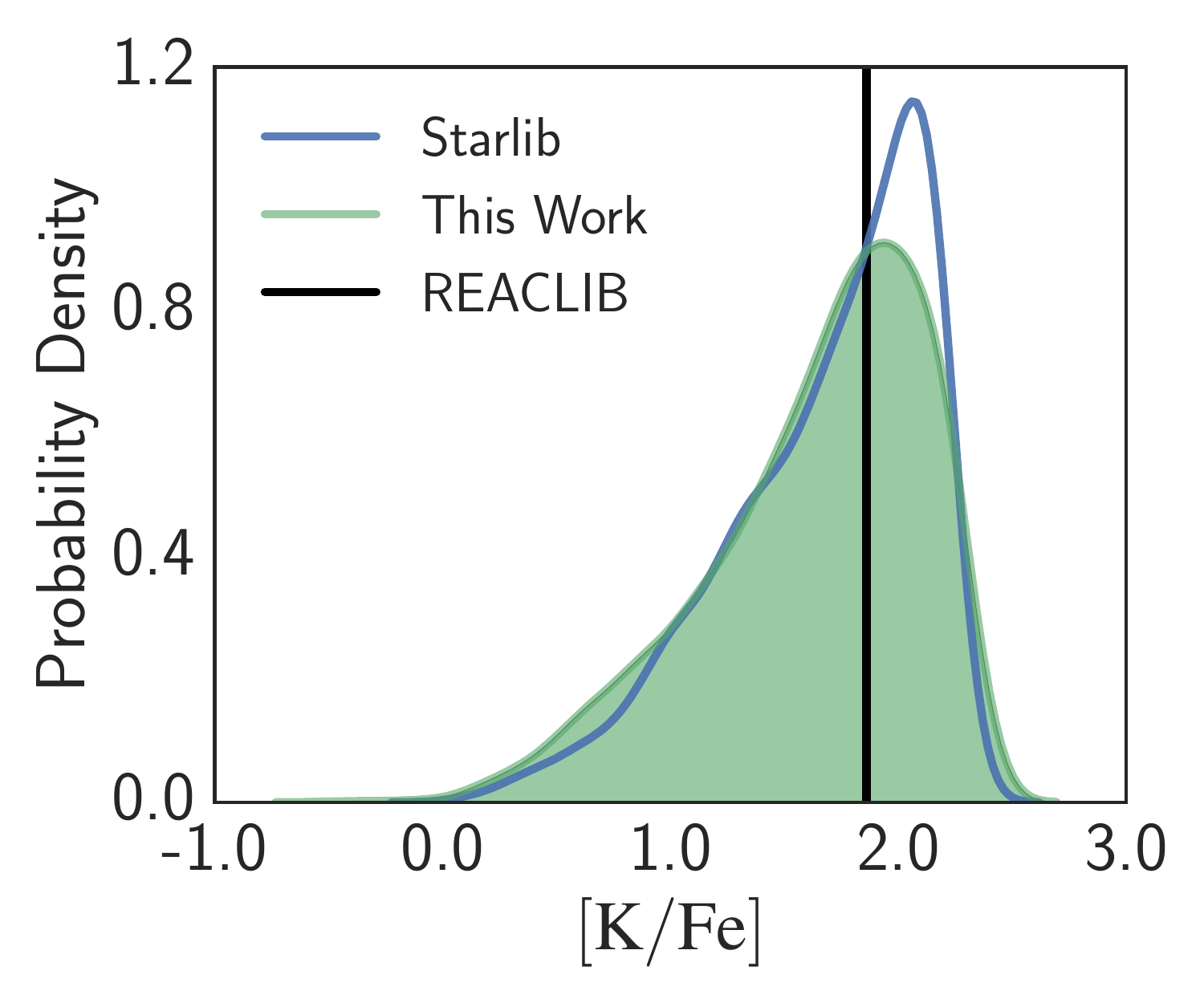}
  \caption{(Color online) Final abundances for potassium obtained for
    the Monte Carlo nucleosynthesis study discussed in the text. A KDE
    was found for the STARLIB rates and our reevaluated rates. The REACLIB rate does
    not define a probability distribution, so its predicated potassium abundance is shown by the solid black line.}
  \label{fig:Final_Abundance_Comp_dex}
\end{figure}

An additional step is to assess how these new rates influence the
predicted elemental potassium abundance.  Spectroscopic observations
are sensitive only to elemental potassium, so its production is a key
constraint on any future theoretical work.  Therefore, the isotopes
\nuc{39}{K}, the long lived \nuc{40}{K}, and \nuc{41}{K} all
contribute to the final observed potassium abundance,
$[\mathrm{K}/\mathrm{Fe}]$, as do the decays of the radioactive nuclei
\nuc{39}{Cl}, \nuc{39}{Ar}, \nuc{41}{Ar}, \nuc{39}{Ca}, \nuc{41}{Ca},
and \nuc{41}{Sc}.  Using the potassium abundance determination from
each individual calculation, a Kernel Density Estimate (KDE)
\cite{Izenman1991} was constructed.  In addition to the updated and
original STARLIB rates, the commonly used REACLIB library rates were
used~\cite{CYB10}.  The REACLIB rates cannot be used in the same Monte
Carlo framework because they do not represent a
complete probability distribution, so their recommended
values were used to provide a single comparison value for the
potassium abundance.  The predicted observable potassium abundance for
each of these cases is shown in
Fig.~\ref{fig:Final_Abundance_Comp_dex}.  The value,
$[\mathrm{K}/\mathrm{Fe}]$ was found to vary up to $\sim 2$ dex for
both Monte Carlo rates; however, with the updated rates the KDE is not
as sharply peaked, and has a greater density toward lower values. This
effect is due to the increased uncertainty for the new rates, which
contributes to a wider spread in the predicted potassium production.
This reinforces the conclusions reached in the correlation study of
Ref.~\cite{Dermigny2017}: the destruction of potassium via \pg is a
crucial process in stellar burning environments, and measurements
aimed at reducing its uncertainty are a necessary step in the study of
the Mg-K anticorrelation in NGC 2419.

\section{Conclusions}
\label{sec:conclusions}

The \pg reaction has been found previously to affect potassium
synthesis in stellar environments leading to the Mg-K anticorrelation
in the globular cluster NGC 2419~\cite{Dermigny_2017}. That finding
was based on estimates of the current experimental uncertainty of the
reaction cross sections, which spurred a thorough re-investigation of
the current experimental picture. 

By considering current experimental measurements of narrow resonances
and including full characterization of upper limits on unobserved
resonance strengths, we present here updated estimates of the rate of
the \pg reaction. The former reaction rate uncertainties also include
ambiguities between experimentally determined resonance strengths
reported in Refs.~\cite{Leenhouts1966}, \cite{Cheng1981},
and~\cite{Kikstra1990}. Correlations between measurements are also
taken into account. The results of this investigation show that the
uncertainties in the \pg reaction are larger than previously
estimated.

The nucleosynthesis ramifications of these findings are also presented
by considering an astrophysical scenario within the bounds established
in Ref.~\cite{Dermigny_2017}. We find that the increased uncertainty
in the \pg reaction rate establishes a clear correlation between it
and the final abundance of \nuc{39}{K}.  Furthermore, the predicted
uncertainty in the elemental abundance of potassium is broadened
towards lower values. Clearly, the \pg reaction must be better
measured if astrophysical scenarios explaining the Mg-K
anti-correlation are to be constrained.


\section{Acknowledgements}
We would like to thank Christian Iliadis and Lori Downen for their
valuable input. This material is based upon work supported by the
U.S. Department of Energy, Office of Science, Office of Nuclear
Physics, under Award Numbers DE-SC0017799 and under Contract
No. DE-FG02-97ER41041.

%

\appendix

\section{Directly Measured Resonance Strengths}
\label{sec:appx-direct}

Direct resonance strength measurements have been performed in the
astrophysical energy region of interest. However, as outlined in
Sec. \ref{sec:resonance-strengths}, several of these measurements are
in disagreement so a Bayesian Maximum Density Posterior Interval
(MDPI) method is employed here to summarize our knowledge of resonance
strengths for Monte Carlo reaction rate calculations. The expectation
value and variance of each resonance strength obtained using this
method is listed in Tab.~\ref{tab:direct-pg}.

{
  \sisetup{
    table-alignment=center,
    table-number-alignment=center,
    retain-zero-exponent=true
  }
  \begin{longtable*}{
      S[round-mode=places,round-precision=1]
      S[table-format=4.2(2)e3]
      S[table-format=4.2(2)e3]
      S[table-format=4.2(2)e3]
      S[table-format=2.2e3]
      S[table-format=2.2e3]
    }
    \caption{\label{tab:direct-pg}Directly measured resonances from
      Kikstra \textit{et al.}~\cite{Kikstra1990}, Cheng \textit{et
        al.}~\cite{Cheng1981}, and Leenhouts \textit{et
        al.}~\cite{Leenhouts1966}. The combined expectation value and
      variance calculated using the method outlined in
      Sec. \ref{sec:resonance-strengths} are shown in the 5th and 6th
      columns}                                                                                                                                                                                                                                              \\
    \toprule \toprule
    & \multicolumn{3}{c}{Literature $\omega \gamma$ (eV)} & \multicolumn{2}{c}{Evaluated $\omega \gamma$ (eV)}                                                                                                                         \\
    \cmidrule(lr){2-4} \cmidrule(lr){5-6}
    {E$_{\text{r}}^{c.m.}$ (keV)} & \multicolumn{1}{c}{Ref.~\cite{Kikstra1990}}         & \multicolumn{1}{c}{Ref.~\cite{Cheng1981}} & \multicolumn{1}{c}{Ref.~\cite{Leenhouts1966}} & \multicolumn{1}{c}{Expect. Val. (eV)} & \multicolumn{1}{c}{$\sqrt{\text{Var.}}$ (eV)} \\
    \midrule
    \endfirsthead
    \multicolumn{5}{l}%
    {{\bfseries \tablename\ \thetable{} -- continued from previous page}}                                                                                                                                                                                   \\
    \toprule
    & \multicolumn{3}{c}{Literature $\omega \gamma$ (eV)} & \multicolumn{2}{c}{Evaluated $\omega \gamma$ (eV)}                                                                                                                         \\
    \cmidrule(lr){2-4} \cmidrule(lr){5-6}
    {E$_{\text{r}}^{c.m.}$ (keV)} & \multicolumn{1}{c}{Ref.~\cite{Kikstra1990}}         & \multicolumn{1}{c}{Ref.~\cite{Cheng1981}} & \multicolumn{1}{c}{Ref.~\cite{Leenhouts1966}} & \multicolumn{1}{c}{Expect. Val. (eV)} & \multicolumn{1}{c}{$\sqrt{\text{Var.}}$ (eV)} \\
    \midrule
    \endhead
    \hline \multicolumn{6}{r}{{Continued on next page}}                                                                                                                                                                                                     \\ \hline
    \endfoot
    \hline \hline
    \endlastfoot
    606.41      & 2.46 (50 )e-2     &                                           &                                               & 2.32e-2                                    & 4.86e-3                           \\
    666.07      & 3.85 (8  )e-2     &                                           &                                               & 3.73e-2                                    & 7.27e-3                           \\
    763.26      & 6.58 (14 )e-2     & 1.38  (41) e-2                            & 1.25 (50 )e-2                                 & 2.78e-2                                    & 4.79e-2                           \\
    807.22      & 1.36 (25 )e-1     & 4.00  (12) e-2                            & 2.5 (10)e-2                                   & 6.15e-2                                    & 1.12e-1                           \\
    881.31      & 8.34 (20 )e-2     &                                           &                                               & 7.85e-2                                    & 1.93e-2                           \\
    898.27      & 5.91 (14 )e-2     &                                           &                                               & 5.60e-2                                    & 1.33e-2                           \\
    1034.06      & 8.28 (21 )e-2     & 1.87  (56) e-2                            & 2.5 (10)e-2                                   & 3.26e-2                                    & 3.99e-2                           \\
    1049.36      & 4.58 (12 )e-2     &                                           & 6.25 (25 )e-3                                 & 2.67e-2                                    & 6.77e-2                           \\
    1059.79      & 4.94 (12 )e-2     &                                           & 1.25 (50 )e-2                                 & 2.70e-2                                    & 3.70e-2                           \\
    1067.30      & 1.70 (50 )e-2     &                                           & 1.25 (50 )e-2                                 & 1.32e-2                                    & 5.67e-3                           \\
    1076.46      & 6.77 (18 )e-2     & 7.25  (22) e-2                            & 8.62 (34 )e-2                                 & 6.75e-2                                    & 2.24e-2                           \\
    1077.92      & 7.52 (12 )e-2     &                                           &                                               & 7.33e-2                                    & 1.22e-2                           \\
    1083.97      & 3.37 (9  )e-2     &                                           &                                               & 3.12e-2                                    & 8.40e-3                           \\
    1090.40      & 1.12 (25 )e-1     & 4.50  (14) e-2                            & 5.75 (23 )e-2                                 & 5.84e-2                                    & 4.14e-2                           \\
    1100.63      & 3.71 (25 )e-2     &                                           & 2.38 (95 )e-2                                 & 2.28e-2                                    & 1.37e-2                           \\
    1104.05      & 4.82 (14 )e-1     & 2.00  (25) e-1                            & 2.5 (10)e-1                                   & 2.60e-1                                    & 1.38e-1                           \\
    1125.49      & 1.46 (38 )e-1     & 5.87  (18) e-2                            & 1.04 (42 )e-1                                 & 8.35e-2                                    & 5.78e-2                           \\
    1171.60      & 7.48 (21 )e-2     & 8.75  (26) e-3                            & 2.5 (10)e-2                                   & 2.99e-2                                    & 5.55e-2                           \\
    1207.96      & 1.92 (50 )e-1     & 5.87  (18) e-2                            & 1.04 (42 )e-1                                 & 9.40e-2                                    & 8.25e-2                           \\
    1209.42      & 4.18 (12 )e-2     & 6.00  (18) e-2                            &                                               & 4.44e-2                                    & 1.74e-2                           \\
    1274.54      & 4.04 (12 )e-1     & 2.12  (42) e-1                            & 1.15 (46 )e-1                                 & 1.72e-1                                    & 1.72e-1                           \\
    1276.19      & 8.41 (25 )e-1     & 5.50  (11) e-1                            & 6.37 (26 )e-1                                 & 5.95e-1                                    & 2.16e-1                           \\
    1312.46      & 8.26 (25 )e-1     & 5.12  (6 ) e-1                            & 6.37 (26 )e-1                                 & 5.84e-1                                    & 1.99e-1                           \\
    1327.18      & 3.61 (11 )e-2     &                                           & 3.75 (15 )e-2                                 & 3.22e-2                                    & 1.23e-2                           \\
    1333.90      & 9.81 (25 )e-2     &                                           &                                               & 9.19e-2                                    & 2.40e-2                           \\
    1340.24      & 3.91 (12 )e-1     & 3.00  (60) e-1                            & 3.75 (15 )e-1                                 & 3.13e-1                                    & 1.03e-1                           \\
    1451.07      & 3.40 (11 )e-1     & 2.25  (45) e-1                            & 3.0 (12)e-1                                   & 2.43e-1                                    & 9.18e-2                           \\
    1456.82      & 1.54 (50 )e-1     & 1.13  (34) e-1                            & 1.04 (42 )e-1                                 & 1.08e-1                                    & 4.54e-2                           \\
    1473.78      & 5.66 (19 )e-2     &                                           & 2.5 (10)e-2                                   & 3.27e-2                                    & 2.55e-2                           \\
    1482.66      & 4.11 (14 )e-2     &                                           & 2.5 (10)e-2                                   & 2.78e-2                                    & 1.51e-2                           \\
    1501.08      & 1.21 (38 )e-1     &                                           & 6.38 (25 )e-2                                 & 7.50e-2                                    & 4.90e-2                           \\
    1506.64      & 9.05 (38 )e-2     &                                           & 3.75 (15 )e-2                                 & 4.90e-2                                    & 4.04e-2                           \\
    1526.13      & 1.64 (50 )e-1     &                                           & 5.00 (20 )e-2                                 & 8.90e-2                                    & 9.97e-2                           \\
    1531.30      & 7.46 (25 )e-2     &                                           &                                               & 6.61e-2                                    & 2.37e-2                           \\
    1536.76      & 8.93 (25 )e-1     & 5.62  (11) e-1                            & 6.88 (28 )e-1                                 & 6.21e-1                                    & 2.37e-1                           \\
    1540.85      & 4.60 (15 )e-1     & 2.63  (52) e-1                            & 3.75 (15 )e-1                                 & 3.07e-1                                    & 1.31e-1                           \\
    1570.19      & 8.80 (25 )e-2     &                                           &                                               & 7.98e-2                                    & 2.40e-2                           \\
    1593.00      & 6.25 (21 )e-2     &                                           & 3.75 (15 )e-2                                 & 4.30e-2                                    & 2.30e-2                           \\
    1611.33      & 1.87 (62 )e-2     &                                           &                                               & 1.69e-2                                    & 5.89e-3                           \\
    1625.56      & 2.29 (75 )e-1     & 8.62  (26) e-2                            & 3.75 (15 )e-1                                 & 1.73e-1                                    & 1.81e-1                           \\
    1648.76      & 1.56 (50 )e-1     &                                           &                                               & 1.40e-1                                    & 4.77e-2                           \\
    1665.23      & 7.05 (25 )e-2     &                                           &                                               & 6.17e-2                                    & 2.35e-2                           \\
    1712.12      & 6.92 (25 )e-2     &                                           & 1.63 (65 )e-1                                 & 8.92e-2                                    & 6.79e-2                           \\
    1720.90      & 6.21 (24 )e-1     & 2.37  (48) e-1                            & 2.37 (95 )e-1                                 & 2.54e-1                                    & 1.78e-1                           \\
    1729.57      & 2.34 (88 )e-2     &                                           &                                               & 2.05e-2                                    & 8.22e-3                           \\
    1752.28      & 1.23 (50 )e-1     &                                           &                                               & 1.04e-1                                    & 4.67e-2                           \\
    1802.29      & 1.87 (75 )e-1     &                                           & 6.88 (28 )e-2                                 & 9.91e-2                                    & 9.00e-2                           \\
    1870.72      & 7.82 (25 )e-2     &                                           &                                               & 7.01e-2                                    & 2.37e-2                           \\
    1876.67      & 3.0 (11)e-2     &                                           &                                               & 2.58e-2                                    & 1.06e-2                           \\
    1882.13      & 1.82 (75 )e-1     &                                           & 7.50 (30 )e-2                                 & 9.57e-2                                    & 8.22e-2                           \\
    1904.35      & 1.67 (62 )e-1     & 7.00  (21) e-2                            & 1.21 (48 )e-1                                 & 9.57e-2                                    & 5.99e-2                           \\
    1934.08      & 1.66 (62 )e-1     & 1.50  (30) e-1                            & 1.38 (55 )e-1                                 & 1.37e-1                                    & 4.67e-2                           \\
    1939.25      & 2.4 (10)  e-2     &                                           &                                               & 2.08e-2                                    & 9.30e-3                           \\
    1946.36      & 3.55 (14 )e-2     &                                           &                                               & 3.07e-2                                    & 1.29e-2                           \\
    1949.48      & 8.87 (38 )e-2     &                                           &                                               & 7.56e-2                                    & 3.48e-2                           \\
    1956.60      & 8.85 (38 )e-2     &                                           &                                               & 7.39e-2                                    & 3.50e-2                           \\
    1990.33      & 1.79 (12 )e0      & 1.79  (12) e0                             & 1.79 (72 )e0                                  & 1.77e0                                     & 2.06e-1                           \\
    2004.17      & 9.95 (38 )e-2     &                                           &                                               & 8.76e-2                                    & 3.53e-2                           \\
    2030.20      & 7.40 (25 )e-2     &                                           &                                               & 6.73e-2                                    & 2.36e-2                           \\
    2033.02      & 2.6 (10)  e-1     &                                           &                                               & 2.23e-1                                    & 9.36e-2                           \\
    2047.06      & 1.47 (62 )e-1     &                                           &                                               & 1.25e-1                                    & 5.87e-2                           \\
    2055.44      & 2.5 (10)  e-1     &                                           &                                               & 2.11e-1                                    & 9.36e-2                           \\
    2086.64      & 7.02 (24 )e-1     &                                           & 5.00 (20 )e-1                                 & 5.16e-1                                    & 2.42e-1                           \\
    2092.29      & 9.66 (38 )e-2     &                                           &                                               & 8.37e-2                                    & 3.52e-2                           \\
    2102.14      & 3.73 (12 )e-1     &                                           &                                               & 3.30e-1                                    & 1.19e-1                           \\
    2112.96      & 3.0 (10)  e-1     &                                           &                                               & 2.67e-1                                    & 9.49e-2                           \\
    2115.49      & 2.04 (62 )e-1     &                                           &                                               & 1.83e-1                                    & 5.95e-2                           \\
    2141.52      & 7.13 (25 )e-2     &                                           &                                               & 6.32e-2                                    & 2.35e-2                           \\
    2150.29      & 1.19 (50 )e-1     &                                           &                                               & 9.97e-2                                    & 4.66e-2                           \\
    2174.66      & 1.29 (50 )e-1     &                                           &                                               & 1.11e-1                                    & 4.71e-2                           \\
    2186.36      & 2.9 (13)  e-1     &                                           &                                               & 2.42e-1                                    & 1.18e-1                           \\
    2199.32      & 4.32 (19 )e-1     & 1.06  (32) e-1                            & 1.50 (60 )e-1                                 & 1.54e-1                                    & 1.37e-1                           \\
    2211.61      & 1.16 (38 )e-1     &                                           &                                               & 1.05e-1                                    & 3.55e-2                           \\
    2223.79      & 2.09 (87 )e-1     &                                           &                                               & 1.78e-1                                    & 8.09e-2                           \\
    2304.31      & 2.4 (10)  e-1     &                                           &                                               & 2.06e-1                                    & 9.20e-2                           \\
    2310.64      & 1.24 (50 )e0      & 6.00  (12) e-1                            & 4.63 (18 )e-1                                 & 5.63e-1                                    & 3.57e-1                           \\
    2317.96      & 1.69 (75 )e-1     &                                           &                                               & 1.43e-1                                    & 6.98e-2                           \\
    2324.78      & 8.99 (38 )e-1     & 1.75  (35) e-1                            & 1.38 (55 )e-1                                 & 2.60e-1                                    & 4.17e-1                           \\
    2341.94      & 2.01 (88 )e0      & 5.75  (12) e-1                            & 9.38 (38 )e-1                                 & 8.51e-1                                    & 5.95e-1                           \\
    2345.25      & 5.59 (25 )e-1     & 8.13  (24) e-2                            & 8.13 (32 )e-2                                 & 1.49e-1                                    & 2.54e-1                           \\
    2368.55      & 3.77 (18 )e-1     &                                           & 1.50 (60 )e-1                                 & 1.91e-1                                    & 1.71e-1                           \\
    2371.08      & 1.11 (50 )e0      & 5.00  (10) e-1                            & 4.12 (16 )e-1                                 & 4.58e-1                                    & 3.02e-1                           \\
    2392.33      & 2.31 (87 )e-1     &                                           &                                               & 2.01e-1                                    & 8.21e-2                           \\
    2409.29      & 5.04 (22 )e-1     &                                           & 1.38 (55 )e-1                                 & 2.45e-1                                    & 3.11e-1                           \\
    2419.33      & 1.64 (75 )e0      & 4.87  (10) e-1                            & 6.88 (28 )e-1                                 & 6.63e-1                                    & 4.49e-1                           \\
    2425.38      & 4.91 (22 )e-1     &                                           &                                               & 4.04e-1                                    & 2.09e-1                           \\
    2441.75      & 7.60 (38 )e-1     &                                           &                                               & 6.18e-1                                    & 3.43e-1                           \\
    2447.90      & 1.74 (75 )e0      & 1.63  (32) e-1                            & 5.00 (20 )e-1                                 & 5.46e-1                                    & 9.17e-1                           \\
    2452.48      & 6.50 (25 )e-1     &                                           &                                               & 5.58e-1                                    & 2.34e-1                           \\
    2459.30      & 3.24 (15 )e-1     &                                           &                                               & 2.65e-1                                    & 1.39e-1                           \\
    2471.58      & 1.19 (50 )e-1     &                                           &                                               & 1.02e-1                                    & 4.61e-2                           \\
    2485.23      & 1.29 (6  )e0      &                                           &                                               & 1.28e0                                     & 6.18e-2                           \\
    2501.61      & 2.9 (13)  e-1     &                                           &                                               & 2.44e-1                                    & 1.15e-1                           \\
    2520.03      & 4.68 (21 )e-1     &                                           &                                               & 3.83e-1                                    & 1.97e-1                           \\
    2540.40      & 5.50 (24 )e-1     &                                           &                                               & 4.61e-1                                    & 2.22e-1                           \\
    2581.54      & 7.32 (38 )e-1     &                                           & 2.5 (10)e-1                                   & 3.48e-1                                    & 3.49e-1                           \\
    2592.65      & 9.38 (50 )e-1     &                                           & 7.00 (28 )e-1                                 & 6.40e-1                                    & 3.55e-1                           \\
    2606.01      & 5.19 (25 )e-1     &                                           &                                               & 4.30e-1                                    & 2.29e-1                           \\
    2623.07      & 1.65 (50 )e0      &                                           &                                               & 1.49e0                                     & 4.75e-1                           \\
    2627.55      & 4.13 (20 )e-1     &                                           &                                               & 3.40e-1                                    & 1.82e-1                           \\
    2647.83      & 9.24 (38 )e-1     &                                           &                                               & 8.03e-1                                    & 3.47e-1                           \\
    2659.52      & 8.19 (38 )e-1     &                                           &                                               & 6.76e-1                                    & 3.49e-1                           \\
    2666.25      & 1.12 (50 )e0      &                                           &                                               & 9.48e-1                                    & 4.63e-1                           \\
    2673.95      & 3.0 (15)  e-1     &                                           &                                               & 2.48e-1                                    & 1.36e-1                           \\
    2682.53      & 1.42 (62 )e0      &                                           &                                               & 1.19e0                                     & 5.81e-1                           \\
    2695.40      & 6.08 (25 )e-1     &                                           &                                               & 5.23e-1                                    & 2.33e-1                           \\
    2713.53      & 6.06 (25 )e-1     &                                           &                                               & 5.13e-1                                    & 2.33e-1                           \\
    2742.19      & 3.1 (15)  e0      & 6.37  (13) e-1                            & 6.37 (26 )e-1                                 & 8.33e-1                                    & 9.79e-1                           \\
    2788.68      & 4.95 (25 )e-1     &                                           &                                               & 3.99e-1                                    & 2.30e-1                           \\
    2798.72      & 5.93 (25 )e-1     &                                           &                                               & 5.06e-1                                    & 2.32e-1                           \\
    2836.84      & 2.0 (10)  e-1     &                                           &                                               & 1.55e-1                                    & 9.19e-2                           \\
  \end{longtable*}
}

\end{document}